# DIRECT DETECTION AND CHARACTERIZATION OF M-DWARF PLANETS USING LIGHT ECHOES

Sparks, William B.[1], White, Richard L.[1], Lupu, Roxana E.[2], Ford, Holland C.[3]

[1] Space Telescope Science Institute, 3700 San Martin Drive, Baltimore, MD 21218, USA.

[2] BAER Institute/NASA Ames Research Center, Moffett Field, CA 94043

[3] The Johns Hopkins University, Baltimore, MD 21218, USA.

Short title: DETECTING EXOPLANETS WITH LIGHT ECHOES

Corresponding author: sparks@stsci.edu

## ABSTRACT

Exoplanets orbiting M dwarf stars are a prime target in the search for life in the Universe. M dwarf stars are active, with powerful flares that could adversely impact prospects for life, though there are counter-arguments. Here, we turn flaring to advantage and describe ways in which it can be used to enhance the detectability of planets, in the absence of transits or a coronagraph, significantly expanding the accessible discovery and characterization space. Flares produce brief bursts of intense luminosity, after which the star dims. Due to the light travel time between the star and planet, the planet receives the high intensity pulse, which it re-emits through scattering (a light echo) or intrinsic emission when the star is much fainter, thereby increasing the planet's detectability. The planet's light echo emission can potentially be discriminated from that of the host star by means of a time delay, Doppler shift, spatial shift, and polarization, each of which can improve the contrast of the planet to the star. Scattered light can reveal the albedo spectrum of the planet to within a size scale factor, and is likely to be polarized. Intrinsic emission mechanisms include fluorescent pumping of multiple molecular hydrogen and neutral oxygen lines by intense Ly$\alpha$ and Ly$\beta$ flare emission, recombination radiation of ionized and photodissociated species, and atmospheric processes such as terrestrial upper atmosphere airglow and near infrared hydroxyl emission. We discuss the feasibility of detecting light echoes and find that under favorable circumstances, echo detection is possible.

1. INTRODUCTION

The study of extrasolar planets and in-depth characterization of those planets, ultimately to determine if any host life, is one of the most important problems facing astronomers today. Recently, planets hosted by late type M dwarfs have come to the fore. Such planetary systems have many advantages for observational studies: the lower mass and radius of the star increase the amplitudes of both transit and radial velocity signals for a given type of planet, and offer the prospect of planetary atmospheric probes for suitable transits (e.g. Benneke & Seager 2012; Kreidberg et al. 2014). The lower luminosities of M dwarf stars move the “habitable zones” for surface liquid water (Kasting et al. 1993) much closer to the stars, which reduces the orbital periods for planets of interest from ~1 year for main sequence G stars to ~10 days for M dwarf stars. Not only that, M dwarfs are by far the most numerous and long-lived stars in the Galaxy, and there are large numbers in the Solar neighborhood. Leveraging these advantages has led to the discovery of roughly Earth-like planets in the habitable zones of Proxima Centauri, the nearest star to the Sun (Anglada-Escude et al. 2016), TRAPPIST-1 (Gillon et al. 2017), and LHS 1140 (Dittmann et al. 2017). TRAPPIST-1 has multiple members of its (currently) seven planet system in the habitable zone. However, despite these advantages, the small angular separation of habitable planets from host M dwarf stars, and the faintness of those stars make them very difficult targets for direct imaging and characterization (Scalo et al. 2007), while transit characterization is feasible only for the small subset of planets that do, in fact, transit their host stars. The geometric probability of habitable planet transits is low (though higher than for G star systems), at ~2% (Charbonneau & Deming 2007), and the stochastic noise associated with active stars may make transit discovery difficult (Parke Loyd & France 2014).

The possibility of life on these nearby planets has generated excitement among both the astronomical community and the public. However, multiple papers have pointed out that powerful stellar flares on these otherwise faint red stars may present major challenges to the development of life, although there are counter-arguments that the planet's atmosphere can survive both the flare events and the associated intense X-rays (Scalo et al. 2007; Zulaga 2013; Bolmont 2017). In addition to the hazards of flares, planets with insolation levels equivalent to Earth orbiting M stars are also likely to be trapped by tidal damping into synchronous rotation, further complicating the actual habitability of these planets (Kasting et al. 1993). Weak magnetic fields resulting from synchronous rotation make planets in M-star habitable zones vulnerable to atmospheric erosion or complete loss caused by coronal mass ejections (Lammer et al. 2007; Khodachenko et al. 2007).

We propose to turn the tables and use the flares as a probe, taking advantage of the short, bright impulse of light from the flare to enhance the brightness of the planet relative to the star, as also suggested independently[1] by Argyle (1974) and Bromley (1992). For example, Proxima Cen, Figs. 1 & 2, has long been known to be a flare star (Shapley 1954), and TRAPPIST-1 also has numerous flares (Vida et al. 2017). Although flares may put strong constraints on habitability, those same, frequent, strong flares, provide an opportunity to enhance the detectability of light from the planet. The rapid variations during flares make the planet easier to detect in the immediate aftermath of the flare, potentially in directly reflected light (Argyle 1974; Bromley 1992), or resonance scattering (Bourrier et al. 2017b), but also exploiting the possibility of a multitude of $H_2$ fluorescence lines pumped by Lyα or FUV continuum, and neutral oxygen fluorescence (1302 Å, 1034 Å, 1306 Å) by Lyβ. The overall battering of the atmosphere by the flaring spectrum from the far ultraviolet (FUV) to X-rays could lead to UV emission through recombination of ionized species such as C II 1334.53 and 1335.71 (France et al. 2010), or enhanced airglow phenomena throughout the electromagnetic spectrum. All of these serve to increase the brightness of the planet relative to the star, and are recognizably distinct from the star due to their time delay, velocity offset and spatial shift, as we discuss below. Scattered light in addition is likely to be polarized which offers a further discriminant of echo light from the host's photospheric starlight.

[1] We encountered these early papers late in the development of the current paper, after initial submission. Both discuss only time series aspects of flare detection. While we were not directly influenced by these works, we acknowledge that they precede us in laying out some of the ideas in this paper.

In the immediate aftermath of a strong flare (due to light travel time delay), the incident flux on a planet, and hence its reflected flux, can be tens to thousands of times its time average. Empirically, flares have been seen with $\Delta V \approx -9$ relative to the star's quiescent flux (Schmidt et al. 2016) and the UV flux can increase by factors $10^4$ to $10^5$ (Reid & Hawley 2005). In favorable circumstances, these factors represent the potential improvement in contrast ratio of the planet to the host star, and while we emphasize the detectability of habitable Earth-like planets, other planets such as super Earths or much larger Jovian sized planets would be correspondingly much easier to detect.

In §2, we briefly review the properties of M dwarf stars and their flares, in §3 we outline fiducial atmospheric models for reference, in §4 we discuss the idealized methodology of using transient brightening and light echoes, §5 considers technical issues and practicalities, including some of the caveats and complexities likely to be encountered. In §6 we examine related situations (disks, the ISM) and §7 discusses the work and presents conclusions.

## 2. M DWARFS AND M DWARF FLARES

### 2.1 The stars

A red dwarf star is a faint main sequence star, defined spectroscopically according to the TiO bands, and M dwarfs, those with absolute visual magnitude $M_V$>7.5 (Reid & Hawley 2005). They range in mass from ≈0.075 M$_\odot$ to ≈0.6 M$_\odot$, and luminosity $10^{-3}$ to 0.07 L$_\odot$, with surface temperatures ≈2100–3800 K. Their photometric properties are presented in Bessell (1991). Reid & Hawley (2005) provide an extended discussion of cool star properties, including M dwarfs and the relevant classification schemes.

While there are many advantages for initial detection of exoplanets around M dwarfs, as described in §1, they present major disadvantages as potential targets for direct imaging and characterization. The angular separation θ of a "habitable" planet from the star is very small. Ignoring spectral modifications to the habitable zone definition, the angular size of the habitable zone is a function of only the apparent bolometric magnitude $m_b$, i.e. the total energy emitted by star as seen from Earth. For example, a Solar-type star at 10 pc with $V$=4.83 has θ=0.1 arcsec. Hence, if we could observe systems with θ > 40 mas, the diffraction limit of the *Hubble Space Telescope* (*HST*) at ~500 nm as a fiducial, we would require $m_b < 7$, for which there are only a handful of M dwarf stars (Scalo et al. 2007, their Fig. 4). Currently available ground-based coronagraphs, and the WFIRST space-based coronagraphic instrument, strive for an inner working angle of only ~0.1 arcsec. Hence even for the nearest few M dwarf systems, the angular separation of potentially habitable planets from their host star presents a very difficult observational challenge for direct imaging and characterization, and for the remainder it is likely to be infeasible for the foreseeable future. Even large space-based mission concepts with coronagraphs or starshades for high contrast starlight suppression, such as the *Habitable Exoplanet Imaging Mission* (*HabEx)* or *Large UltraViolet Optical InfraRed Surveyor (LUVOIR)* will only be able to directly image and characterize a small number of M dwarf exoplanet systems. (A small percentage of such exoplanets, ~2%, will transit their host star, enabling transit-style characterization for those, as discussed elsewhere.)

In the absence of bolometric corrections, the contrast ratio of planet brightness to star brightness would scale inversely with the stellar luminosity for a planet of given size and albedo receiving the same degree of insolation. For example, if the contrast ratio of the Earth to the Sun is $C_1 \approx 10^{-10}$, then for an equivalent planet around a star with 1% of the Solar luminosity the contrast ratio would be $C_2 = 10^{-8}$. However, in the case of M dwarfs, much of the bolometric luminosity is in the infrared, and the resulting observed contrast ratio is $C_2 = 10^{-8}(L^*/L^*_{\rm bol})$ where $L^*$ is the red dwarf luminosity in the observed band, and $L^*_{\rm bol}$ its bolometric luminosity. Hence, while the contrast ratio of the planet to star is improved by proximity of the planet to the star, this advantage is mitigated by the bolometric ratio (the stars are very red), and the small angular separation. Other factors can come into play to improve the contrast ratio of a potentially habitable planet, such as albedo, orbital phase and size (which in the case of Proxima Cen b is a function of the unknown orbital inclination). Proxima Cen b is expected to have a contrast ratio $\approx 10^{-5}$–$10^{-7}$ (Turbet et al. 2016; Kane et al. 2017, as well as our own calculations in §3.).

With light echoes, we avoid a bolometric correction and improve the baseline contrast ratio by a factor corresponding to the star's luminosity during a flare compared to outside a flare, which can be orders of magnitude. We are not necessarily restricted to flaring wavelengths, as will be discussed below, where we consider fluorescence and enhanced airglow at non-flaring wavelengths.

### 2.2 The flares

While M dwarfs are long lived, small, faint stars, they are prone to dramatic, rapid brightening, particularly in the blue, U-band and ultraviolet wavelengths, in the form of stellar flares analogous to Solar flares. The fraction of M dwarf stars that are active is highest from M4 to M9 (Reid & Hawley 2005), where the stars become fully convective. In seconds, the star can increase in brightness by orders of magnitude. This rapid rise is followed by a rapid decline, the "impulsive phase" (Bopp & Moffett 1973; Kowalski et al. 2013), and a longer timescale "gradual decay phase". The impulsive phase rise and fall has a typical timescale of ~tens of seconds, and is the most relevant for triggering light echo events, Fig. 2 (additional examples are shown later in §5, Fig. 9). The decay phase can last minutes or hours (Kowalski et al. 2013). Kowalski et al. (2013)

present an atlas of 20 dMe flare events, and define an impulsive index *I* which is the ratio of the flare flux over its quiescent value, to the flare timescale, indicating the dominance of the impulsive phase to the gradual decay, $I = I_{f,\mathrm{peak}}/t_{1/2}$ where $t_{1/2}$ is the full width half-maximum of the flare time series. Of the twenty flares studied, 11 are "impulsive" and four more "hybrid", with contrast enhancement factors ranging from a few to 208×. The U-band energy release covers a wide range, reaching $7.9\times10^{33}$erg in a large flare on AD Leo. Pettersen et al. (1986) show that the distribution of decay times is strongly peaked towards $t_{1/2} < 20$ sec (their Fig. 9), with a number of unresolved flares of duration < 9 sec.

Parke Loyd & France (2014) present an extensive discussion of the UV spectral characteristics of flares, and find strong flares in Si IV, Si III and the continuum for Proxima Cen, AD Leo and others in the context of the stochastic variability of the stars and our ability to detect transits.

Kowalski et al. (2016) present a comprehensive analysis of the blue and near ultraviolet (NUV) continuum properties of dMe flares (see also Reid & Hawley 2005). The spectral ingredients are typically a strong blue continuum of temperature ~$10^4$ K or hotter well-described by a black-body spectrum, and plentiful associated line emission, especially in the FUV which is dominated by strong chromospheric and transition region line emission (Fig. 1).

In addition to the intense burst of optical and UV radiation which is characteristic of flares, as in Solar flares, there is likely to be significant expulsion of energetic particles in the form of coronal mass ejections (CMEs), and emission of high energy EUV and X-ray photons. While not strictly light echoes, these all may impact and influence the exoplanet presumed to orbit the active M dwarf star over a range of timescales, see §5.3.

## 3. STEADY STATE

For comparison purposes, we include a brief estimate of the thermal and scattered light for one of these planets in the habitable zones of M-stars, namely Proxima Cen b. Such models provide a baseline for time domain observations.

The largest uncertainty for the atmospheric models is the composition. This affects the entire thermodynamic equilibrium in the atmosphere, and is reflected in the spectral signatures in both thermal emission and reflected light. Bixel et al. (2017) conclude that there is a 90% probability that Proxima Cen b is a rocky planet, and a 10% probability that it is a mini-Neptune with a H/He envelope. For the purposes of illustration, we investigate three possible compositions:

a. Earth-like, with a total surface pressure of 1 bar, and a composition of 80% $N_2$, 18% $O_2$, 1% $CO_2$, and 1% $H_2O$.
b. Venus-like, with extra water, with a total surface pressure of 90 bar, and a composition of 10% $N_2$, 80% $CO_2$, and 10% $H_2O$.
c. Neptune-like, with a total surface pressure of 200 bar, and a composition determined by equilibrium chemistry for a metallicity of +1.7 dex and a solar C/O ratio. This results roughly in 76% $H_2$, 4%$H_2O$, and 2.5% $CH_4$.

In addition, we consider a "Hot Earth" model, with the same composition and additional internal heating, as an example of how our approach may be generalized. Other planet parameters are kept the same among models: $a$ = 0.0485 AU is the semi-major axis, and $g$ = 14.4 m $s^{-2}$ is the surface gravity (Anglada-Escudé et al., 2016).

Proxima Cen, the illuminating star, is an M6 star with an effective temperature of ~3050 K, a radius of 0.141 $R_{\odot}$, a specific gravity log($g$) = 5.2, and characterized by strong activity. We reproduce these characteristics by using a BT-NextGen model from the Phoenix database (Allard et al. 2012), with $T_{\rm eff}$ = 3100 K, and log($g$) = 5, extended to the far-UV using a model for the Early Sun from Claire et al. (2012), to simulate the expected activity level..

We derive the atmospheric pressure-temperature profiles and the thermal emission (Fig. 3) using a well-tested radiative-convective equilibrium code (McKay et al. 1989, Marley et al. 1999, Lupu et al. 2014). We find that all the cold models have effective temperatures between 247.5 and 248.5 K, slightly higher than the predicted equilibrium temperature of 234 K (Anglada-Escudé et al., 2016) and in agreement with their positioning in the habitable zone, while the hot model has an effective temperature of 693.5 K. However, these models also show that, depending on composition and total atmospheric mass, the surface temperatures are 259 K, 523 K, and 1750 K for the three cold models, respectively, making some of these scenarios likely incompatible with life. As seen in Fig. 3, all cold models cross the water condensation curves (dashed lines), and therefore water clouds will form at these elevations. Such an insight is important for further estimating the scattering properties of these atmospheres.

Further, we estimate the planet's albedo, and its contrast ratio in reflected light, using a custom version of the code described in Marley et al. (1999) and Cahoy et al. (2010) (with K. Feng and T. Robinson, paper in preparation). Fig. 4 shows the reflected light spectra for the three cold planet scenarios, as a function of phase angle. For a rough comparison, we estimate the reflectivity in the presence of clouds by assuming a bright, optically thick cloud deck at the altitude corresponding to water condensation for each case, respectively, but keeping the pressure-temperature profile unchanged. A hot planet, such as our fourth scenario, would be easier to detect in thermal emission, and would be likely devoid of water clouds, and was excluded from this comparison.

A brief stellar flare will likely not change the equilibrium pressure-temperature-profiles and the effective radiating temperatures of these planets. However, the reflectance spectra point out the favorable observing windows where stellar activity will be less obscured by atmospheric absorption, as well as the variation of reflectivity with phase angle. Other intrinsic transient emission from the planet will arise on top of these equilibrium features.

## 4. TRANSIENT SIGNATURES AND IDEALIZED METHODOLOGY

When a flare occurs on the surface of the star in a location visible to the planet, there will be a corresponding flare from the planet. That flare may simply be reflected light, or it may be enhanced by fluorescence and other forms of intrinsic emission in the atmosphere due to the incident burst of broad-band irradiation from the X-rays to the UV, and the strong stellar flare emission lines of Lyα and Lyβ. There may also be a mechanical contribution from the flare outflow as it encounters the planet (Segura et al. 2010; Luger et al. 2017). While the Lyman lines of the star, which pump fluorescence, can be strongly attenuated from our perspective by transmission through the interstellar medium (France et al. 2012; France et al. 2013; Linsky et al. 2013), they are not attenuated from the perspective of the planet.

The light from the planet can be distinguished from the light from the star in several ways, with the relevant equations summarized in Table 1 and geometry illustrated in Fig. 5. For both scattered and emitted light, we can discriminate by means of the time delay, the velocity shift, and positional shift. For scattered light, we can potentially also distinguish the echo from the star by polarization of the scattered light, and intrinsic or stimulated emission, by the unique (non-stellar) character of the emission and its correlation with flaring activity. Due to geometric effects, we may not always see the same flares as the planet. This issue presents challenges and complexities but also opportunities, which we discuss in §5.4.

*(1) Time delay*

The size of the time delay between the flare and observations of its arrival at the planet has a predictable variation with orbital phase, with an amplitude that depends on the orbital inclination and distance of the planet from the star (orbital radius). Depending on the phase of the orbit and inclination of the orbital plane, the time delay as seen from Earth for the echo signal from the planet can be up to twice the light travel time from the star to the planet (when the planet is at superior conjunction behind the star for an inclination near 90°).

*(2) Velocity shift*

The motion of the planet in its orbit induces a Doppler shift in light scattered by the planet. For example, with a semi-major axis of 0.0485 AU and a period of 11.186 days, the circular orbital velocity of Proxima Cen b is 47 km $s^{-1}$. Emission lines from the planet, either due to reflection of emission lines of the flares or to planetary atmospheric emission that is excited by the energy input from the flare, will be shifted by up to $\pm 47 \sin i$ km $s^{-1}$, where $i$ is the inclination of the orbit. This is well within the capability of spectrographs with moderately high resolution, R~20,000, Fig. 2.

*(3) Position shift*

The light from the planet will be spatially shifted compared with the position of the star. While for many M dwarfs, this will be a very challenging measurement, as discussed in §1, for a few it offers the prospect of an additional echo discriminant. The astrometric offset of the Proxima Cen b planet depends on the phase and inclination of the orbit, but at the greatest elongation is about 0.04 to 0.06 arcsec (see Fig. 2 of Kane et al. 2017). Positional precision to these levels is certainly possible with *HST*, and may be possible with ground-based adaptive optics systems in favorable circumstances. A positional offset will change in tandem with both the time delay and velocity shift, Table 1.

*(4) Polarization*

Scattered light is polarized to a greater or lesser degree. Different types of scattering lead to a variety of diagnostics. By measuring the planet's polarization as it orbits the star, we obtain a segment of the polarization phase function, the span of the segment depending on the orbital inclination. For an orbit of inclination $i$ to the line of sight, the range of accessible scattering angles is 90-$i$ to 90+$i$ degrees, which always includes 90° scattering, where polarization is likely to be high. The polarized spectrum contains diagnostics that differ from spectroscopy alone (Stam et al. 2004; Stam 2008; Zugger et al. 2010). Features in the polarization phase function can reveal the presence of liquid droplets, clouds and their composition, via rainbows (Bailey 2007; Karalidi et al. 2012) and ocean glint as the scattering angle passes through the Brewster's angle. For water, the Brewster's angle is ≈53°, which results in a highly polarized specular reflection, hence the maximum polarization degree for ocean glint occurs at phase angle $\Phi=\cos^{-1}(-0.276/\sin i)$ which is 106° for an edge-on system. The inclination needs to exceed 16° for the polarization phase function to include the Brewster's angle for water. Generically, scattered light polarization characteristics differ for clouds, hazes, solid surfaces (rocks, ices), and liquids. In extreme, the polarization spectrum can yield information on the presence of life, either through linear polarization enhancing surface biosignatures (Berdyugina et al. 2016), or through the circular polarization spectrum which may serve as a remote sensing diagnostic of homochirality, one of the most promising universal biosignatures (Wald 1957; Sparks et al. 2009, 2015; Jafarpour et al. 2015, 2017).

*(5) Intrinsic emission: resonance scattering, fluorescence and airglow*

The presence of an exoplanet atmosphere or extended exosphere may be revealed by resonance scattering of strong line emission, potentially enhanced during strong flares. A Lyα reconnaissance of the TRAPPIST-1 system has been carried out by Bourrier et al. (2017a) who detected marginal flux decreases during transits, hinting at possible hydrogen exospheres around the inner planets. Stronger variations were found on Kepler-444, which are consistent with neutral hydrogen exospheres trailing the two outer planets of that system (Bourrier et al. 2017b). While these studies depend on the transit configuration, analysis of resonantly scattered line emission in the context of flaring frees us from that requirement, and vastly expands the number of systems amenable to study, utilizing not only Lyα, but other strong lines such as CIV, CII, etc., Fig. 1. The combination of a time-delayed response with orbital Doppler shift, or time-modulated Doppler shift, would be strong evidence of a planetary origin.

We can seek evidence of additional intrinsic emission in the exoplanet atmosphere by exploiting the possibility of a multitude of $H_2$ fluorescence lines pumped by Lyα, and neutral oxygen fluorescence (1302 Å, 1034 Å, 1306) by Lyβ. The Lyα, Lyβ flux is likely to be much higher viewed by the planet, 5–20× due to the absence of interstellar attenuation, and hence fluorescent emission lines can potentially arise in the exoplanet atmosphere. The velocity shift between the pumping line from the star (e.g. Lyα) and the absorbing species in the planet's atmosphere can also be revealed by the strength of the emission lines, even when a radial component of the velocity towards the observer cannot be measured. Similarly, the powerful impulsive energy input may result in strongly enhanced airglow lines and potential electron-impact excitation and aurorae (Luger et al. 2017). Intrinsic line emission may be recognized by both its physical nature, i.e. can the molecule be present in the stellar photosphere, and by a velocity shift which depends on the phase angle and inclination.

*(6) Putting it all together*

The two most robust parameters to unambiguously distinguish a planetary echo from stellar emission are the time delay Δt, and the radial (line of sight) component of orbital velocity *v*. These two parameters do not change with increasing distance of the star from Earth, and are closely coupled by Kepler's Laws. They follow a well-defined relationship which depends on the orbital inclination. Fig. 6 shows the loci of Δt versus *v* as a function of inclination for circular orbits. Departures from these curves gives insight into non-circular or non-Keplerian motion, introduced by orbital ellipticity, Fig. 7, or the presence of dynamically interacting companions. The expectation is that habitable zone M-dwarf planets are likely to be tidally locked into circular orbits due to tidal dissipation, however the empirical upper limit to the eccentricity for Proxima Cen b is only loosely constrained at *e*=0.35. The full equations are given in Tables 1 and 2, while the extrema of velocity, time delay and positional offset for circular orbits are given by:

$$\Delta t_{max} = 49.9\left(\frac{r}{0.05\,AU}\right)\frac{(1+\sin i)}{2} \text{ seconds} \tag{1}$$

$$v_{max} = 42.1 \sin i \sqrt{\left(\frac{M_*}{0.1 M_\odot}\right)\left(\frac{0.05\,AU}{r}\right)} \text{ km s}^{-1} \tag{2}$$

$$v_{max} = 42.1 \sin i \sqrt{\left(\frac{M_*}{0.1 M_\odot}\right)\left(\frac{49.9s}{\Delta t}\right)\frac{(1+\sin i)}{2}} \text{ km s}^{-1} \tag{3}$$

$$\Delta x_{max} = 50\left(\frac{r}{0.05\,AU}\right)\left(\frac{1}{D\,pc}\right) \text{ milliarcsec} \tag{4}$$

where $M_*$ is the mass of the star, and *r* is the radius of the planet's orbit around the star and *D* is the distance of the star from Earth. Here the parameters have been scaled to the (approximate) nominal properties of the Proxima Cen planet, Proxima Cen b, (Anglada-Escude et al. 2016). Note that velocity can be both positive and negative, hence the differential velocity shift between the approaching and receding configurations is $2v_{max}$.

It is important to note that in the case of a circular orbit of known period, a *single* measurement of the time delay and velocity shift is sufficient to provide the inclination and hence mass of the planet (Fig. 6). A lower limit to the size is available by assuming the albedo is unity, with assumptions about the phase function. While (to a good approximation) circular orbits are perhaps to be expected, in the event that orbits are elliptical, the orbital inclination, ellipticity and angle of periastron can be determined by measuring Δt versus *v* as a function of time, Fig. 7.

The extraction of an echo signal from the data is discussed in Appendix A. If both time delays and velocity shifts are available, a maximum likelihood algorithm based on the Richardson-Lucy iteration can be used to extract both the stellar light curve and the planet's contribution. In the event that we only have a time series, without velocity, positional, or polarimetric corroborating observations, or if the flare duration is longer than the light travel time to the planet, we can employ more sophisticated mathematical methods to identify the presence of a faint echo of the primary light curve, delayed by the light travel time to the planet. The most straightforward approach is probably auto- or cross-correlation in the time dimension (Argyle 1974; Sparks & Ford 2002; Snellen et al. 2013, 2015). An extension of this algorithmic approach could be an autocorrelation with a variable time lag that depends on the orbital phase. The physics of flares may lead to "bounces" in the aftermath of a flare that resemble echoes, so structure in a single flare would not suffice to prove the detection of a planet. However, with a sufficient number of flares plus the smoothly changing delay predicted at a range of phases, it should be possible to identify echoes using purely photometric time series data. The short flare from the star produces a distinctive pulse shape that can be used as a filter in searching for the planet signal, which in turn allows stacking the light curves from many flares to search for the planet's signal. With a vast literature on the analysis of time series and the identification of echoes, coupled to enormous commercial interest (audio, sonar, ultrasound, seismic exploration, for example), we anticipate this will be fertile ground for future dedicated research.

## 5. OBSERVATIONAL STRATEGIES AND FEASIBILITY

We examine three wavelength regimes: FUV, NUV/optical and optical/NIR, and discuss their response to the presence of strong, fast flares on the host star. There could be a thermal IR response, however we would not expect a thermal response time to be fast, though Meng et al. (2016) detect a 4.5 μm emission time lag due to variable incident stellar radiation in reverberation mapping of a protoplanetary disk around a T Tauri star. There could be stimulated radio emission from the flare interaction with a planetary magnetosphere, and X-rays from high energy atmospheric or surface processes. We do not discuss these other regions of the electromagnetic spectrum, though we recognize that each offers a unique window into potential detection and characterization of exoplanets. We make reference to existing instrumentation to understand the orders of magnitude involved, and anticipate possible optimized dedicated instrumentation capabilities, as well as offering generic descriptions of each wavelength regime.

### 5.1. Far ultraviolet (FUV)

The FUV spectra of M dwarf flares is replete with chromospheric and transition region emission lines, while photospheric continuum emission is very low (e.g. France et al. 2012; Fig 1). During a flare, weak FUV continuum emission is seen (France et al. 2012). From a technical perspective, it is not uncommon for FUV data to be obtained using "time tag" mode, in which individual photon arrival events are detected and stored in a table, with their position and time of arrival as entries in the table. On board *HST*, both the *Space Telescope Imaging Spectrograph* (*STIS*) and *Cosmic Origins Spectrograph* (*COS*) can be used in this way. This is an extremely versatile observing mode, allowing the time resolution of the data products to be adjusted across a very wide range, and individual features to be isolated in windows of time, wavelength and space. The two dimensional detectors of *STIS* can encode either position, in direct imaging, or wavelength, as in the echelle spectrum of Fig. 1. Hence specific emission lines can be studied, individual flares, or groups of either, or in direct imaging, a set of filters allows selection of spectral windows. Here, we focus on spectroscopy.

i. Reflected emission line light

Fig. 1 shows a two dimensional echelle spectroscopic dataset obtained with *HST STIS* of the FUV spectrum of Proxima Cen. A large number of chromospheric and transition region emission lines are visible and all flare dramatically, Fig. 2. Echo emission of reflected flares would be subject to a time delay and velocity delay, related through the equations presented in Tables 1 and 2, and discussed in §4. The intensity of reflected light is expected to be given by the normal wavelength-dependent contrast ratio, as shown in Fig. 3, but *relative to the flare brightness rather than the quiescent stellar brightness*. Fig. 3 shows that the FUV/NUV albedo can depend strongly on atmospheric composition, cloud properties, and phase angle. In these models the FUV/NUV spectrum is dominated by Rayleigh and Raman scattering, but molecular and atomic features can impose additional albedo structure, depending on the details of the composition.

To get a rough idea of whether echo detection may be feasible, Hawley et al. (2003) and Osten (2017) find a correlation between the CIV flux and *U* band flux during the impulsive phase of AD Leo flares, with $\log f_{CIV} = 0.32 + 0.92 \log f_U$, $f$ in erg s$^{-1}$ cm$^{-2}$ for the line and erg s$^{-1}$ cm$^{-2}$ Å$^{-1}$ for the continuum. If this relation holds for other stars and energetic flares, then for a large flare of energy $E_U$=10$^{33}$ erg (Appendix B) say from Proxima Cen released over the impulsive phase in 10 sec, for a telescope of effective area 1000 cm$^2$ (e.g. a 1-m telescope of 13% efficiency), a total of ≈7×10$^6$ CIV photons would be detectable from the star, ~70 from a planet with contrast ratio 10$^{-5}$, hence, since the planet echo photons are displaced in time and wavelength from those of the star, a signal-to-noise (*S/N*) of *S/N*~8 results from a single event. Clearly much more detailed analysis is required to assess ultimate feasibility, however we do note that CIV is not the strongest line (Si IV and Lyα are significantly stronger, with Lyα ≈37× the CIV flux, Osten 2017) and there are a large number of additional FUV lines, with very low continuum. We find these numbers encouraging and sufficient to warrant additional study.

ii. Fluorescent emission lines

Molecular $H_2$ lines are seen in the M dwarf spectra of the *Measurements of the Ultraviolet Spectral Characteristics of Low-mass Exoplanet Host Stars* (MUSCLES) survey of weakly active M dwarfs (France et al. 2012; France et al. 2013), and are readily detected in the spectrum of Proxima Cen, Fig. 1. To date, due to the lack of an observed velocity shift, these lines have been attributed to emission from the stellar photosphere (France et al. 2013). France et al. (2013) acknowledge that additional modeling is needed to explore the intriguing possibility that there is an exoplanet contribution to the fluorescence. Interestingly these fluorescent lines are not observed in M-stars that are not known to host planets while, all four of the stars in France et al. (2013) showing strong $H_2$ fluorescence host Neptune to super-Jupiter mass planets (France et al. 2013). The O I triplet at 1302 Å also fluoresces when excited by Lyβ emission. In detail, there are significant complexities to the analysis and prediction of the origin and location of fluorescent species, and these will be presented for Proxima Cen elsewhere (paper in prep.[2]).

---

[2] A particular challenge for the existing *STIS* data for Proxima Cen is that, by unlucky chance, the orbital phase of the observations puts emission from the planet near zero velocity, at the same

In addition, the flare energy absorbed by the atmosphere can give rise to a chain of dissociations and ionizations that will in turn have their own spectral signatures and characteristic time scale. For example, electron-impact excitation of molecular hydrogen is one of the most prominent features in the FUV spectrum of Jupiter (e.g. Yelle & Miller 2004). The characteristic feature of this process lies between 1550 and 1620 Å, and its detection would be potentially feasible at a lower spectral resolution than required for detailed analysis of fluorescent emission.

### 5.2 Near ultraviolet (NUV)

The NUV spectra of flares contain both a black body thermal contribution, and strong line emission (Hawley et al. 2007). For light echo observations there are advantages to both continuum and line emission. Continuum emission allows sampling of the planetary albedo spectrum and is more likely to lend itself to polarization observations. Line emission offers the prospect of Doppler shifted echo discrimination, bringing with it improved contrast ratio as the echo light is shifted away from the light of the host star not only in time, but also in wavelength, and improved ability to recognize an echo as such (Appendix A). Observed NUV flare spectra exhibit both emission lines and continuum, with comparable amounts of energy in each. The strongest emission lines are MgII and a complex of Fe II emission lines. The continuum is relatively stronger in the more energetic flares (Hawley at al. 2007). Kowalski et al. (2016) derived a temperature of 12,100 K.

The Swift Ultraviolet/Optical Telescope (SWIFT) broad-band NUV light curve for Proxima Cen is shown in Fig. 8, see also Appendix B. There are numerous strong, fast flares. Fig. 9 shows detailed profiles for a sample of UV flares. Fig. 10 shows that the strongest flares in this sample even have the shortest duration. Others have also found that flare duration timescale does not correlate (or does so only weakly) with flare amplitude (Güdel et al. 2003) which is helpful from the light echo perspective.

To assess the *S/N* from echo observations, we consider three scenarios (i) integrating the planet and star together (ii) seeking light echoes after each suitable flare using the time averaged flare luminosity and (iii) seeking light echoes after the brightest flare(s) only. If we define the *S/N* to be the number of collected photons from the planet divided by the square root of the number of photons collected from the star during the flare, with the appropriate time delay, we have (see Appendix B for details):

$$S/N \approx C\sqrt{L_* t} \propto t^{1/2} \quad (5)$$

$$S/N \approx b\sqrt{d}C\sqrt{L_* t} \propto t \quad (6)$$

$$S/N \approx b_{max}C\sqrt{L_* t_f} \propto t^{3/2} \quad (7)$$

where $C$ is the contrast ratio of the planet, $L_*$ is the quiescent stellar luminosity, $b$ the average amplification factor during flares (i.e. the star is $b\times$ brighter during a flare), $t$ is time and $d$ is the duty cycle (the fraction of time the star is flaring). The approximate scaling relations for *S/N* with time $t$ are derived using the expressions in Appendix B and are appropriate for the flare

---

wavelength as local geocoronal line emission. The O I line has a red component that might have been attributed to the planet, but on closer examination is seen to vary with the Hubble orbital period and almost certainly has a terrestrial origin.

distributions for Proxima Cen or AD Leo. (We determined that instrumental and background noise contributions are negligible for COS, and LUMOS according to its design requirements, France et al. 2017. The noise is dominated by the quiescent stellar flux for the duration of the flare, which is very much higher than the background equivalent flux of LUMOS, Fig. 9 of France et al. 2017, Appendix B.)

Note that the maximum energy flare has $E_{\max} \propto t^{1.5}$ (approximately, equation A1a) and the time averaged luminosity in flares $L_{flares} \propto \sqrt{t}$ (approximately, equation A2). This has the interesting effect that the *time averaged (flare) brightness of the star increases with time*. The noise $N$ is given by the stellar quiescent emission, *but only that which is accumulated during the flaring*, which we parameterize by $d$, the duty cycle. In the worst case (from an $S/N$ perspective), if all the emission is continuum, the average amplification factor $b$ during a flare is proportional to the time averaged flare luminosity, hence $S/N \propto t$ (approximately). Hence, although identification of flare echoes will be a difficult observation, longer duration observations have an advantage with $S/N$ increasing with $t$ rather than the usual $\sqrt{t}$ for the average of all flares seen.

To underscore the effect of time, we consider the echo from a single, very luminous flare of brightness $b_{max}$ and duration $t_f$. Since $b_{max} \propto E_{\max}$ for a fixed flare duration, we find $S/N \propto t^{1.5}$ (approximately), provided that the flare energy distributions continue to high energy. As $E_{\max}$ increases, we are free, post facto, to reject weaker flares, correspondingly diminishing the noise from the background quiescent stellar emission. This dependence, $S/N \propto t^{1.5}$, is even faster than that of the time average, above, and quantitatively offers the theoretical best $S/N$.

To obtain a handle on the orders of magnitude involved, we consider three fiducial scenarios. *COS* on *HST* with the G230L grating covers the spectral regions 170–210 and 280–320 nm simultaneously with an effective area of 585 $cm^2$ at 260 nm (Fox et al. 2017). The decadal survey 15-m mission concept study *LUVOIR*[3], with the *LUVOIR MultiObject Spectrograph (LUMOS)* instrument concept (France et al. 2017) provides an effective area $\approx 1.8 \times 10^5 cm^2$ for the G300M grating, a factor ≈300× improvement over *HST* in the NUV[4]. A hypothetical 1-m space telescope optimized for the NUV with system efficiency 15% (compared to 13% for *LUMOS*) would have an effective area ≈2× that of *COS*.

Bromley (1992) estimated the detectability of light echoes from photon statistics alone, and concluded that a single observation with a 1-m telescope could detect a Jupiter-size planet in an orbit of radius less than 0.1 AU around AD Leo and a planet the size of Neptune would be observable if its orbital radius were less than 0.05 AU. Using our time averaged analysis, for Proxima Cen (see Appendix B), and a planet with contrast ratio $C=10^{-5}$, the time taken to reach $S/N=3$ without additional strategies to enhance the contrast (such as Doppler shift), would be 43 days for *COS*, 21 days for the hypothetical 1-m, and 2 days for *LUMOS*. For the brightest flare method, equation (7), the time to reach $S/N=10$ would be 9.4 days (*COS*), 5.7 days (1-m) and 1.2 days (*LUMOS*). For less favorable contrast ratios, the averaged flare approach can reach $C \approx 10^{-6}$ in 232 days, for the 1-m, but the brightest flare method accomplishes this (in an idealized world) in only 11 days, and in principle can go to $C \approx 10^{-7}$ in ~ 55 days integration for $S/N=3$, or 130 days for

[3] https://asd.gsfc.nasa.gov/luvoir/

[4] See also http://jt-astro.science:5102/lumos_etc for a preliminary exposure time calculator for *LUVOIR/LUMOS*

*S/N*=10 and C≈$10^{-7}$. Hence even Earth/super Earth sized planets are potentially within range with observations of ~days to ~weeks.

This is highly idealized of course, and assumes all the flare energy is available to drive the echo albeit all in continuum. If we take into account a typical flare time profile, about half the energy is in the impulsive phase, and half in the gradual decay phase. Then, in the *S/N* formulae above, equations 5–7, we have offsetting effects – the planet counts diminish by ≈2×, the contaminating stellar flux countrate increases by ≈2× due to residual emission from the decaying flare, but the time over which it is collected decreases by ≈10×. The net results is *S/N'*≈*S/N*×0.5/✓(2×0.1)≈1.1, a net increase in *S/N* by 11%. Subtleties at this level will require detailed modelling to truly sort out, and are beyond the scope of the present paper whose intent is to focus on a conceptual analysis of the light echo phenomenon.

The steep dependence of *S/N* on time dominates over aperture, especially in the case where we look at the brightest flares. The required integration times can be shortened by introducing additional contrast enhancement (Doppler shifted emission lines), although these are likely to be offset by the complications and practicalities discussed in §5.4 and elsewhere. The scaling relations tend to favor a long-term, smaller dedicated "staring" mission, which would allow post-facto windowing and selection of data. The method of equation (5) would be better attempted in the optical or NIR where there is plentiful light from the star, and other methods such as planet-star Doppler shift, spectral structure and polarization, can be brought to bear (Charbonneau et al 1998; Collier Cameron et al. 2002; Sparks & Ford 2002; Snellen et al. 2013, 2015; Schworer & Tuthill 2015).

There are ≈80 stars with NUV fluxes within a factor 10 of Proxima Cen's in the HAbitable Zones and M dwarf Activity across Time (HAZMAT) program of Shkolnik & Barman (2014), chosen from young moving groups and nearby old stars, indicating that from even this incomplete sample, there is a useful number of observable stars on the sky. Dressing & Charbonneau (2015) further estimated that there are ≈ 0.45 Earth and super Earth's per M dwarf habitable zone, suggesting that amongst the observable stars, there exists a large number of undetected exoplanets in the habitable zones, potential targets for light echo methods used as a tool for discovery.

### 5.3 Visible/NIR airglow lines

There are a variety of physical mechanisms and timescales associated with the onset and evolution of a flare. We have discussed the impulsive phase UV radiation, primarily, but there are also the optical and UV emissions of the gradual decay phase, X-ray and EUV emission, and energetic particle release in the form of CMEs and other geomagnetic disturbances in the stellar wind.

Each of these has a characteristic timescale. The impulsive phase lasts ~seconds to tens of seconds, the gradual decay phase ~1000 sec, while X-rays and EUV rise and fall over ~3000 s to a few hours (Hawley et al. 1995; Güdel et al. 2002; Güdel et al. 2003). The X-ray emission profile is qualitatively similar to the time derivative of the optical, the "Neupert effect" (Hawley et al. 1995; Güdel et al. 2002). A geomagnetic disturbance in a 400 km $s^{-1}$ stellar wind would take 5.2 hrs to traverse 0.05 AU, while a fast CME event travelling at 2500 km $s^{-1}$ (Khodachenko et al. 2007) would arrive in 50 min. The amount of radiant energy in each of the optical/UV, EUV and X-ray bands is comparable, in the range ~$10^{29}$ to $10^{34}$ erg (Hawley et al. 1995; Audard et al. 2000; Güdel et al 2003). These extended timescales offer the possibility of a protracted atmospheric response.

Segura et al. (2010) consider the impact of both high energy electromagnetic radiation (UV and X-ray) and associated proton flux (energies >10MeV) from flares on the atmospheric chemistry of an Earth-like atmosphere. While there is an immediate response to the impulsive phase of a flare, the atmosphere evolves more slowly afterwards, returning to its equilibrium after ~$10^6$–$10^7$sec. Luger et al. (2017) calculate the strength of possible auroral emission from Proxima Cen b, caused by a stellar wind with magnetospheric outbursts from Proxima Cen. The inferred strength of the 577.7 nm OI auroral line is not detectable with current instrumentation, but with future instruments and telescopes it could become so.

Khodachenko et al. (2007) and Lammer et al. (2007) discuss the likely impact of CMEs on planets around M dwarfs. Luger et al. (2017) discuss the enhancement of auroral emission due to these and similar perturbations and conclude that auroral emission excited by energetic particles is likely to dominate over airglow excited by EUV and X-rays. Nevertheless, if high energy photons are incident at ~$3\times10^{30}$erg $s^{-1}$ ($10^{34}$ erg over 1 hr), for an Earth-radius planet at 0.05 AU the incident energy is ~135 TW. The fraction converted to airglow depends completely on the photochemistry of the upper atmosphere. If there is significant ozone present, essentially all the light shortward of 320 nm will be absorbed in the ozone Hartley band, and potentially be available for atmospheric chemistry and excitation processes, such as excitation of the rotational-vibrational Meinel series of hydroxyl (OH) emission lines in the near infrared.

Leinert et al. (1998) tabulated the strongest terrestrial airglow lines, and the OH line series is by far the strongest, at 4.5 MR (mega-Rayleighs, where R is the Rayleigh unit of surface brightness), compared to, e.g., the green oxygen line at 558 nm of 250 R, or 10 kR auroral 558 nm emission for typical zenith airglow brightnesses.

With the high energy photon flux, and stellar wind and CME timescales being comparable, high resolution spectroscopic searches for optical/NIR line emission are likely to benefit from filtering according to this timescale – the optimal times to seek an atmospheric response are during and for the few hours following strong flares. An instrument such as the *CRyogenic high-resolution InfraRed Echelle Spectrograph* (CRIRES) on the European Southern Observatory's *Very Large Telescope* (VLT) would seem well-suited to a search of this nature, or the *High Accuracy Radial velocity Planet Searcher* (HARPS) spectrograph analyzed in detail by Luger et al. (2017). Lovis et al. (2017) discuss coupling the high resolution spectrograph ESPRESSO to the adaptive optics instrument SPHERE at the VLT in order to measure both a spatial offset and a Doppler shift for Proxima Cen b, and conclude that such a measurement is feasible. If additional filtering on the flares were introduced, signal-to-noise can potentially be improved further.

### 5.4 Complexities and opportunities

The geometric distribution of flares on the surface of the star is unknown. Hunt-Walker et al. (2012) found no correlation with rotational phase for AD Leo, similar to Hawley et al. (2014), indicating a distribution of many flare sources at different longitudes and a relatively uniform flare distribution, stochastic in both outburst frequency and location. A non-uniform flare distribution could influence the illumination pattern of the space local to the star, and either work in favor if flares are concentrated to the equatorial regions, or against if to the poles (presuming the planets to share the same equatorial plane).

The planet may be illuminated by flares which we do not see, because they are both “around the back” of the star, and conversely, we may see flares which the planets do not. The degree to which flares are visible is also not strictly known. A conservative assumption would be for a flare visibility over $2\pi$ steradians (i.e. if the flare is on the visible hemisphere). However, flares are extremely luminous compared to the star, and may occur, at least in part, at significant height above the photosphere (e.g. Hawley et al. 2007). Hence flares may be visible over a more than a $2\pi$ sr hemisphere, due to scattering or other stellar atmospheric responses or if there is significant flare emission at altitude. For a hypothetical planetary system, we always (potentially) see the planet when it is illuminated by a flare, but we may not see the same flares as the planet. For a face-on system, and $2\pi$ sr visibility, we see half the flares the planet sees, and the planet sees half the flares we see. For an edge-on system, when the planet is behind the plane of the sky, we see none of the flares it sees, but all of them when it is in front. Compare this to a traditional estimate of completeness – the fraction of systems with habitable planets that transit the habitable zone of M dwarfs is ~1.5–2.7% (Charbonneau & Deming 2007) and the remaining 98% never transit. Hence characterization by transit observations is never possible, due to geometry, for ≈98% of systems. For light echoes however, at least in principle, *all* systems which flare are potentially amenable to study.

If there are longer duration bright flares invisible to us but illuminating the planet, recalling that bright flares offer the best signal to noise characteristics, we may use all the reflected light and not just the initial impulsive phase. Can we recognize the echo of a powerful flare, in isolation? If it is individually detectable, which statistics suggest can be the case, we may be able to recognize it through Doppler shift, fluorescent emission, polarization, or perhaps through some other distinctive characteristic. For example, it is possible that the strongest flares may exhibit unique spectral features, or have a recognizable shape to the light curve. In such a circumstance, the complication is our ability to recognize the echo, but the opportunity is that we may utilize the entirety of the flare emission, including the gradual decay phase.

With intermittent data, determined by the stochasticity of flare occurrence, establishing periodicities may prove challenging. Typical habitable zone planets will have periods ~tens of days, however the strongest flares occur less frequently, while the signal accumulation time needed to utilize weaker flares can also be lengthy. Establishing periods in irregularly sampled data is a classical problem, but important to resolve in this context.

Flares may have significant intrinsic velocity relative to the star: ultraviolet emission in Solar flares is seen blueshifted, with redshifted recombination hydrogen emission (Druett et al. 2016). Since in general we expect to see both the flare and echo spectra, this ought not be a major issue. While some examples have been found of asymmetries in M dwarf flare velocity profiles (e.g. Gunn et al. 1994; Hawley et al. 2003), overall the velocity distribution is generally symmetric (Reid & Hawley 2005), with both blue and red enhancements and relatively small velocities ~4 km $s^{-1}$ (Hawley et al. 2007).

## 6 DISKS, CLOUDS AND THE INTERSTELLAR MEDIUM

For anticipated coronagraphic observations designed for exoplanet detection and characterization, the presence of an exozodiacal disk can be a major problem. Clumps in the disk can mimic planets, and diffuse emission can contribute noise and limit detection capability. The three dimensional geometry of a light echo is an expanding ellipsoid with the observer at one focus, and the source (flare star) at the other – to a good approximation, a paraboloid with focus at the star. For light echoes, the presence of an exozodiacal disk in the system is significantly less relevant. The extreme brevity of M dwarf flares means that the fraction of a disk illuminated by a flare is small (the intersection of the disk and the expanding paraboloid surface of the echo whose width is the distance covered at light-speed during the flare), unlike in a conventional direct imaging observation where the whole disk is illuminated all the time. Coupled to a typically low intrinsic optical depth for a disk, in most cases we do not expect to see light echoes from disks.

There may be individual examples of flaring stars imbedded in regions of sufficiently high dust density to allow detection of light echoes from scattering dust, and given typically longer timescales for light propagation into a disk, we would be able to use flares of longer duration and flares hidden behind the host star (Gaidos 1994). The well-known M star AU Mic hosts a gas-poor dusty debris disk, and Plavchan et al. (2009) discovered several additional debris disks around low mass stars. Meng et al. (2016) used Spitzer observations and “photo-reverberation” 4.5 μm mapping to determine the essential geometric properties of a protoplanetary accretion disc around a T Tauri star. Ribas et al. (2017) found a possible infrared excess for Proxima Cen potentially due to a warm ring or disk of dust, while Anglada et al. (2017) detected multiple dust components in the system with ALMA.

In principle, echoes from dusty disks can be used to map the geometry of the disk, and seek gaps, rings and clumps that may be indicative of the presence of planets, as the echo sweeps through the disk. The signatures of such an echo response would be extended duration, polarization, and an evolving Doppler shift as the echo propagates through the dust. Gaidos (1994) discusses the circumstances and disk ages for which light echoes may present a useful probe of flare star disks.

Similar considerations apply if the flaring star is located within a relatively dense, dusty ISM, such as the T-Tauri and Herbig Ae/Be stars whose light echoes were studied by Ortiz et al. (2010). There is a trade between increasing scattering cross-section and flare contrast to the blue and UV, mitigated by higher attenuation of the light. Rapidly expanding forward scattered light echo rings, as in SN1987A (Crotts 1988) are generally easier to observe. At the Taurus molecular cloud, for example, at a distance of ~140 pc, where young, active stars are present, a transverse echo moving at $c$ propagates at 53 mas/hr, well within the capabilities of *HST*’s resolution, while forward scattered faster moving rings can be amenable to ground based observation.

## 7 DISCUSSION AND CONCLUSIONS

While we have concentrated on habitable Earth-like exoplanets, there is a wide range of planet types know to orbit M dwarfs, including an approximately equal number of Earth-sized and super Earth sized ($1.5{-}2\ R_\oplus$) planets (Dressing & Charbonneau 2015). Super Earths with ~2–4× the area of an Earth-sized planet, all other things being equal, will be brighter by the same factor. Planets the size of Jupiter are known to exist in orbit around M stars (Hartman et al 2015), and will be much brighter still, have much more favorable contrast ratios and will be relatively easily observed, both within and somewhat beyond the habitable zone. For fast enough flares, it is possible to probe the space interior to the habitable zone, where contrast ratio is further improved for a given planet. Planets of all types are ultimately of interest to the community, and the search by means of light echoes offers a new way to, literally, illuminate this entire class of objects.

Exoplanets are very faint in an absolute sense, however there are considerations that can enhance a light echo in certain situations. These include: ***(i) inclination*** of the planet's orbit – in many cases the inclination is unknown, presuming the planet to have been detected using precision radial velocity techniques. If it is close to face-on, the planet's size (strictly speaking, mass) must increase due to the sin $i$ term in the radial velocity detection, and hence also the strength of the echo signal compared to the case of a higher inclination orbit; ***(ii) fluorescence***, from a plethora of $H_2$ lines and the oxygen triplet pumped by Lyα, Lyβ respectively, lines which are seen in M dwarf spectra could potentially be identified with the planet by using velocity, time delay and position. The Lyα and Lyβ fluxes, which are the relevant parameters for fluorescent excitation will be much higher viewed by the planet than inferred from Earth observations, 5–20×, due to the absence of interstellar attenuation (France et al. 2013; Linsky et al. 2013); ***(iii) the entire flare energy*** from the UV through X-rays is incident upon the planet, as well as an energetic particle flux from related CME and geomagnetic storm activity, not just the thermal continuum, leading to excited emission from the planet's atmosphere following the blast of ionizing radiation and energetic particles (Luger et al. 2017). While the timescales are longer, such processes can contribute to the detectability of the planets; ***(iv) extent of the atmosphere*** of the planet and possibility of outflow of gas from the planet – a large gaseous halo and trailing tail can be an effective scatterer of FUV radiation, should one be present (Bourrier et al. 2017b), which would increase the planet's cross section without affecting the mass estimate; ***(v) the presence of ancillary material*** associated with the planet, such as Saturn-like rings, can increase the reflected flux by a significant factor, while remaining consistent with the mass estimate; ***(vi) additional planets:*** if there are multiple planets in the system (TRAPPIST-1 has seven), they will all create their own light echoes, recognizable by their time delay, velocity, polarization and positional shifts. A repeating sequence of small amplitude blips in a light curve could be a compelling indication that we are witnessing a multiple system such as TRAPPIST-1. Multiple Earth sized planets in compact orbits have also been found by Astudillo-Defru et al. (2017), albeit somewhat interior the habitable zone.

In conclusion, we have outlined ways in which M dwarf exoplanets, emphasizing those in the habitable zone, may potentially be identified by light echoes and/or stimulated emission due to flares on the M dwarf star. The principles carry over to any active star, and do not require the planet to transit in front of the star. The observations are challenging, but offer the prospect of a greatly improved contrast ratio for some of the most interesting exoplanet targets, in both time and spectral domains and may bring into the realm of observation, planets that may never (at least for for the foreseeable future) be observable using coronagraphic techniques even for future large space mission concepts.

Echo emission is recognizable by a time delay due to the light travel time from the star to the planet, by its Doppler shift relative to the star due to the orbital motion of the planet, by a spatial offset of the emission compared to the position of the star in a few cases, and by a polarization signature characteristic of scattered light. Each of these discriminants serves not only to identify echo emission, but also to provide important physical and geometric diagnostics of the planet and its motion around the star. The time delay and Doppler shift are closely coupled by Kepler's laws.

From the perspective of detectability, and recognition of a feature as a light echo, we think it is important that multiple discriminants be brought to bear. In particular, the presence of a significant Doppler shift will be a powerful complement to assessing whether a small intensity enhancement could be an echo, versus an irregular decline in the flare itself.

Light echo methods can potentially bring habitable zone Earths and super Earths into the realm of feasibility for observation, along with larger planets and planets both inside and outside the habitable zone. Well known systems such as Proxima Cen b are tractable to the approach, while current statistics indicate that there are numerous additional potential targets. The most promising wavelength regions to use are in the blue and NUV spectral regions, where the contrast of flare to star is high, and the stellar quiescent emission is relatively weak, and the FUV where the flares are exceptionally fast and strong, the stellar quiescent emission is almost non-existent, and a host of strong emission lines dominates the flux, improving contrast through Doppler shift and time, and there are expected to be fluorescent lines uniquely present in a planetary atmosphere.

We propose that the use of light echoes from M dwarf and other flares star offers a timely, exciting and potentially powerful new probe of some of the most interesting and scientifically important planetary systems that have been identified and which remain to be discovered.

ACKNOWLEDGEMENTS

Some of the data used in this paper were obtained using the Hubble Space Telescope which is operated by STScI/AURA under grant NAS5-26555. We acknowledge the use of public data from the Swift data archive. The *HST* and Swift data presented in this paper were obtained from the Mikulski Archive for Space Telescopes (MAST). STScI is operated by the Association of Universities for Research in Astronomy, Inc., under NASA contract NAS5-26555. Support for MAST for non-*HST* data is provided by the NASA Office of Space Science via grant NNX09AF08G and by other grants and contracts. RL wishes to thank Channon Visscher and Kat Feng for continuing contributions and collaboration. This research was supported by NASA and NSF grants. This research has made use of data and/or software provided by the High Energy Astrophysics Science Archive Research Center (HEASARC), which is a service of the Astrophysics Science Division at NASA/GSFC and the High Energy Astrophysics Division of the Smithsonian Astrophysical Observatory.

APPENDIX A

*Extracting the light echo from data*

To separate the faint signal of a planet from the stellar emission, it is necessary to weight the counts from the planet more heavily at times and wavelengths where the signal-to-noise is highest. The noise is contributed almost entirely by the star. The appropriate weighting is highest in the aftermath of a flare, when the star's emission is relatively faint.

Consider a simplified data analysis problem where the noise is the result of photon-counting statistics and both the time delay $\Delta t$ and the velocity shift $\Delta v$ are assumed to be known. Also assume that the planet's signal is an exact (but reduced in amplitude by a factor $A$) replicate of the star's light curve, shifted by the known amounts in time and velocity. The maximum likelihood solution for $A$ using optimal weighting is

$$A = \sum_{ij} P_{ij} \left( {}^{y_{ij}}\!/_{S_{ij}} - 1 \right) \Big/ \sum_{ij} P_{ij}^2 y_{ij} / S_{ij}^2 \qquad (5)$$

where the summation is over time and wavelength, $S_{ij}$ is the brightness of the star, $P_{ij}$ is the brightness of the planet, and $y_{ij}$ is the observed number of counts in the bin. The planet brightness $P_{ij}$ is assumed to have the same scale as the star, so $\ll 1$ . The uncertainty in $A$ can be then computed using error propagation:

$$\sigma^2(A) = 1 \Big/ \sum_{ij} P_{ij}^2 / S_{ij} \qquad (6)$$

Note that if the planet is not shifted either in wavelength or time, the summation is equal to the number of counts in the star. In that case the variance is 1 over the number of counts, as expected for Poisson statistics. The summation in the denominator can be used to determine a factor by which the signal-to-noise ratio increases for a given set of signals:

$$f = \left[ \sum_{ij} P_{ij}^2 / S_{ij} \Big/ \sum_{ij} S_{ij} \right]^{1/2} \qquad (7)$$

This factor can be used to compare the ability to detect a planet with and without the time delay and wavelength shifts. The signal-to-noise improves by a factor $f$, and the time to reach a given SNR is reduced by a factor $f^2$. If there are no shifts at all, $f = 1$.

If the wavelength profile and brightness versus time are separable, so $S_{ij} = T_i W_j$ , then the sums in $f$ are also separable. The SNR factor is then a product of factors from the velocity shift and the time delay.

As a concrete example, we have calculated the SNR factor for Proxima Cen using the line profile from the STIS E140M observations and the flare shown in Figure 9(d). The $f$ factor for only the wavelength shift is 2.9, for only the time delay is 1.6, and for both effects combined is 4.5. For a stronger flare, the time delay factor increases (but the wavelength shift factor is unchanged.) If the flare is stronger by a factor of 10 (which is still easily within the observed range), the time delay factor increases to 3.2 and the factor for both shifts combined is 9.2. That implies that the planet could be detected with an integration time a factor of 84× smaller by using both the time delay and velocity shift.

APPENDIX B

*Equations for NUV flare statistics*

We use the following to apply the *S/N* expressions of equations (5) – (7), and derive the scaling relations given in §5.2, equations (5)–(7). Let the cumulative flare distribution be described by $\log \nu = m \log E_U + c$ where ν is the rate at which flares with energy in the *U* band greater than $E_U$ are produced by the star, and *m* and *c* are constants which parameterize a particular star. We introduce a factor α which is the energy of the flare in the observational passband relative to that in the *U* band: $E = \alpha E_U$. Then the number of flares with energy greater than *E* in the chosen passband in time *t* is ν*t*.

After a time *t*, the expected maximum energy for a flare is $E_{max}$, given by ν*t*=1:

$$E_{max} = \alpha\, 10^{-c/m} t^{-1/m} \qquad \text{(A1a)}$$

The expected time for a flare of energy $E_{max}$ to occur is:

$$t = \alpha^m 10^{-c} E_{max}^{-m} \qquad \text{(A1b)}$$

Integrating the luminosity weighted frequency distribution, the time averaged luminosity in flares $L_f$ is:

$$L_f = \frac{1}{\beta} \alpha 10^{-c/m} t^{\beta} \qquad \text{(A2)}$$

where $\beta = -(1+m)/m$.

If the star is on average a factor *b* brighter during a flaring episode and the star is flaring for a fraction *d* of the time, then:

$$b = \frac{L_f/d}{L_*} \qquad \text{(A3)}$$

where *d* is the duty cycle and $L_*$ is the quiescent luminosity of the star in the chosen passband.

If the brightest flare lasts for time $t_f$ then the maximum brightening factor $b_{max}$ for the star is

$$b_{max} = \frac{E_{max}/t_f}{L_*} \qquad \text{(A4)}$$

For Proxima Cen, Davenport et al. (2016) used the *Microvariability and Oscillations of STars microsatellite* (MOST; Walker et al. 2003), to find 63 flares per day with 0.5% amplitude, and ~8 flares per year with energy ~$10^{33}$ erg. The corresponding frequency of flares was log ν = -0.68 log $E_{MOST}$+20.9 (Davenport et al. 2016, ν in flares per day). Pettersen et al. (1986) found that for AD Leo, there were 0.66 flares per hour with energies over $10^{30}$ erg in the *U* band over 1971–1985. They also found that for flares above $10^{30}$ erg, the cumulative distribution is approximated by log (*N/T*)=14.7-0.62 log $E_U$, consistent with other work (Lacy et al. 1976; Hunt-Walker et al. 2012), where *N* is the number of flares with energy larger than $E_U$ (erg) and *T* is in seconds.

Hence, the corresponding values for the slope *m* are -0.68, -0.62 and for the intercept *c*, 15.84 and 14.7 with ν in flares/second, for Proxima Cen and AD Leo respectively. Inserting these numbers into equations (A1)–(A4) yield the scaling relations given in the main text, §5.2, equations (5)–(7). That is, $m \approx -0.65$, $\beta \approx 0.5$ . If we assume $L_*$, *d*, $t_f$, $\alpha, \beta$, *c*, *m* are constant in time, then for equation (6) which described the *S/N* for a time averaged analysis (the signal from all useable flares is averaged), the number of collected photons from the planet is $N_p = CL_* bdt$ and the number of photons collected from the star is $N_* = L_* dt$ hence $S/N = N_p/\sqrt{N_*} = Cb\sqrt{L_* td}$, and from equations A2 and A3, we see:

$$S/N\,(t) \propto b\sqrt{t} \propto L_f\sqrt{t} \propto t^{\beta+0.5} \propto t \qquad \text{(approximately)} \qquad \text{(A5)}$$

For equation (7) which is the echo *S/N* from the single brightest flare, the number of collected photons from the planet is $N_p(max) = CL_* b_{max} t_f$ and the number of photons collected from the star is $N_* = L_* t_f$ hence $S/N = N_p(max)/\sqrt{N_*} = Cb_{max}\sqrt{L_* t_f}$, and from equations A4, A1a, we find:

$$S/N\,(t) \propto b_{max} \propto E_{max} \propto t^{-1/m} \propto t^{1.5} \quad \text{(approximately)} \qquad \text{(A6)}$$

For $E_U \approx 0.66 E_{MOST}$, i.e. $\alpha_{MOST} \approx 1.5$, the Proxima Cen time averaged luminosity in flares is $L_{Uf} \approx 5\times10^{26}$ erg s$^{-1}$ for a maximum flare luminosity of $E_U$(max)=$10^{33}$erg reached after ~45 days, a factor of two to ten less than AD Leo. Hawley et al. (2014) found for one of the most active stars in the Kepler field, GJ1243 (dM4e), that $L_{Uf} \approx 3\times10^{28}$ erg s$^{-1}$.

For a black body of temperature ≈12,100K (Kowalski et al. 2016), the energy emitted in the range 170–320 nm is ≈3.8× the energy in the *U* band (Bessel 2005), hence $\alpha_{NUV} \approx 3.8$ with lower stellar contamination than in the *U* band. Such a black body has a *total* energy ≈7–8× the energy within the *U* band.

SWIFT's UltraViolet Optical Telescope (UVOT) (Roming et al. 2005) observed Proxima Cen repeatedly between 2009 and 2016. In total it acquired 24 hours of time-tagged photometry in the *uvw1* filter (central wavelength 260 nm, FWHM 69.3 nm). The light curve is presented in Fig. 8. Numerous strong, rapid flares are apparent, consistent with the rate of Davenport et al. (2016) using the photometric zeropoint of Breeveld et al. (2011). About 1.5% of the time the star is undergoing a flare more than 20-σ brighter than the quiescent emission in this NUV band. The implied quiescent magnitude of Proxima Cen at 260 nm is AB=16.8.

## REFERENCES

Allard F., Homeier, D., Freytag, B, 2012, RSPTA 370, 2765

Anglada, G., Amado, P.L., Ortiz, J.L. et al. 2017, ApJL submitted, astro-ph/1711.00578

Anglada-Escude, G., et al. 2016, Nature 536, 437.

Argyle, E. 1974, Icarus, 21, 199.

Astudillo-Defru, N., Díaz, R.F., Bonfils, X. et al. 2017, A&A, 605, L11.

Audard, M., Güdel, M., Drake, J.J., Kashyap, V.L., 2000, ApJ, 541, 396.

Bailey, J., 2007, Astrobiology, 7, 320.

Benneke, B., Seager, S. 2012, ApJ, 753, 100.

Berdyugina, S.,V., Kuhn, J.R., Harrington, D.M., Šantl-Temkiv, T.; Messersmith, E. J., 2016, IJAsB, 15, 45.

Bessell, M.S. 1991 AJ, 101, 662.

Bessel, M.S., 2005, ARAA, 43, 293.

Bixel, A, Apai, D., 2017, ApJ, 836, L31.

Bolmont, E. et al. 2017, MNRAS 464, 3728.

Bourrier, V., Ehrenreich, D., Wheatleey, P.J., et al., 2017(a), A&A, 599, 3.

Bourrier, V., Ehrenreich, D., Allart, R., et al., 2017(b), A&A, 602, 106.

Breeveld, A.A., Landsman, W. Holland, S.T. et al. 2011, in Gamma Ray Bursts 2010. AIP Conference Proceedings, 1358, 373.

Bromley, B. 1992, PASP, 104, 1049.

Cahoy, K.L. et al. 2010, ApJ, 724, 189.

Charbonneau, D., Jha, S., Noyes, R.W. 1998, ApJ, 507, L153.

Charbonneau, D., Deming, D. 2007, submitted to the Exoplanet Task Force (AAAC), 2 April 2007, arXiv:0706.1047.

Claire, M.W., et al., 2012, ApJ, 757, 95.

Collier Cameron, A., Horne, K., Penny, A., & Leigh, C. 2002, MNRAS, 330,187.

Crotts, A., 1988, ApJ, 333, 51.

Davenport, J.R.A., Kipping, D.M., Sasselov, D., Matthews, J.M., Cameron, C. 2016, ApJ, 829, L31.

Dittmann, J.A., Irwin, J.M., Charbonneau, D. et al. 2017, Nature, 544, 333.

Dressing, C.D., Charbonneau, D., 2015, ApJ, 807, 45.

Druett, M. Scullion, E., Zharkova, V., Matthews, S., Zharkov, S., Van der Voort, L.R. 2016, Nature Communications, DOI: 10.1038/ncomms15905

Fox, A. J., et al. 2017, *Cosmic Origins Spectrograph Instrument Handbook*, Version 9.0 (Baltimore: STScI)

France, K., Stocke, J., Yang, H.  et al., 2010, ApJ, 712, 1277.

France, K., Linsky, J.L., Tian, F., Froning, C.S., Roberge, A. 2012 ApJL, 750, L32.

France, K., Froning, C., Linsky, J., Roberge, A., et al. 2013, ApJ, 763, 149. astro-ph/1212.4833.

France, L., Fleming. B, West, G. et al. 2017 Proc SPIE 2017; 10397-39, astro-ph/1709.06141.

Gaidos, E. 1994, Icarus, 109, 382.

Gillon, M., et al. 2017, Nature 542, 456.

Güdel, M., Audard, M., Kashyap, V.L., Drake, J.J., Guinan, E.F. 2003 ApJ 582, 423.

Gunn, A.G., Doyle, J.G., Mathioudakis, M., Houdebine, E.R., Avgoloupis, S. 1994, A&A, 285, 489.

Hartman, J.D., Bayliss, D., Brahm, R. et al. 2015, AJ, 149, 166. doi:10.1088/0004-6256/149/5/166

Hawley, S.L., Fisher, G.H., Simon, T., Cully, S.L. et al., 1995, ApJ, 453, 464.

Hawley, S.L., Allre, J.C., Johns-Krull, C. et al. 2003, ApJ, 597, 535.

Hawley, S.L., 2007, Walkowicz, L.M., Allred, J.C., Valenti, J.A. PASP, 119, 67.

Hawley, S.L., Davenport, J.R.A., Kowalski, A.F., Wisniewski, J.P., Hebb, L., Russel, D., Hilton, E.J. 2014, ApJ, 797, 121.

Hunt-Walker, N.M., Hilton, E.J., Kowalski, A.F., Hawley, S.L., Mattews, J.M. 2012, PASP, 124, 545.

Jafarpour, F., Biancalani, T., Goldenfeld, N., 2015 Phys. Rev. Lett. 115, 158101.

Jafarpour, F., Biancalani, T., Goldenfeld, N., 2017 Phys. Rev, 95, 032407.

Judge, P.G., Kleint, L., Sainz Dalda, A., 2015, ApJ, 814, 100.

Kane, S.R., Geling, D.M. & Turnbull, M.C. 2017, AJ 153, 52.

Karalidi, T., Stam, D.M., Hovenier, J.W. 2012, A&A, 548, 90.

Kasting, J.E., Whitmire, D.P., Reynolds, R.T., 1993, Icarus, 101, 108.

Khodachenko, M.L., Ribas, I., Lammer, H., Grießmeier, J.-M., et al., 2007, *Astrobiology* 7, 167.

Kreidberg, L., Bean, J.L., Désert, J.-M., et al. 2014, Nature, 505, 69.

Kowalski, A.F., Hawley, S.L., Wisniewski, J.P. et al. 2013, ApJS, 207, 15.

Kowalski, A.F., Mathioudakis, M., Hawley, S.L., Wisniewski, J.P., Dhillon, V.S., Marsh, T.R., Hilton, E.J., Brown, B.P. 2016, ApJ, 820, 95.

Lacy, C. H., Moffett, T. J., & Evans, D. S. 1976, ApJS, 30, 85.

Lammer, H., Lichtenegger, H.I.M., Kulikov, Y.N. et al. 2007, Astrobiology, 7, 185.

Leinert, Ch., Bowyer, S., Haikala, L.K. et al., 1998, A&ASupp, 127, 1.

Linsky, J.L., France, K., Ayres, T. 2013, ApJ, 766, 69.

Lovis, C., Snellen, I., Mouillet, D. et al. 2017, A&A 599, 16.

Luger, R., Lustig-Yaeger, J., Fleming, D., Tilley, M.A., Agol, E., Meadows, V.S., Deitrick, R., Barnes, R. 2017, ApJ, 837, 631.

Lupu, R.E. et al., 2014, ApJ, 784, 27.

Marley, M.S., McKay, C.P., 1999, Icarus, 138, 268.

McKay, C.P. et al., 1989, Icarus, 80, 23.

Meng, H.Y.A., Plavchan, P. Rieke, G.H., Cody, A.M. et al., 2016, ApJ, 823, 58.

Ortiz, J.L., Sugerman, B.E.K., de la Cueva, I., Santos-Sanz, P., Duffard, R., Gil-Hutton, R., Melita, M., Morales, N. 2010, A&A, 519, A7. DOI: 10.1051/0004-6361/201014438

Osten, R. 2017 Instrument Science Report STIS 2017-02; STScI

Parke Loyd, R.O., France, K. 2014, ApJS, 211, 9.

Pettersen, B.R., Panov, K.P., Sandmann, W.H., Ivanona, M.S. 1986, A&AS, 66, 235.

Plavchan, P., Werner, M.W., Chen, C.H., Stapelfeldt, K.R., So, K.Y.L., Stauffer, J.R., Song, I., 2009, ApJ, 698, 1068.

Reid, I.N., Hawley, S.L., 2005, New Light on Dark Stars: Red Dwarfs, Low-mass Stars, Brown Stars, Praxis, Springer-Verlag., ISBN 3-540-25124-3.

Ribas, I., Gregg, M.D., Boyajian, T.S., Bolmont, E. 2017, A&A, 603, 58.

Roming, P. W. A., Kennedy, T. E., Mason, K. O., et al. 2005, Space Sci. Rev., 120, 95.

Scalo, J., Kaltenegger, L., Segura, A., Fridlund, M. et al. 2007, Astrobiology, 7, 85

Schmidt, S.J.; Shappee, B.J.; Stanek, K. Z. ; Prieto, J.L.; Holoien, T.W.-S.; Kochanek, C.S., 2016, in The 19th Cambridge Workshop on Cool Stars, Stellar Systems, and the Sun (CS19), Uppsala, Sweden, 06-10 June 2016, id.71; doi: 10.5281/zenodo.57865

Schworer, G., Tuthill, P.G. 2015, A&A, 578, 59.

Segura, A., Walkowicz, L, Meadows, V., Kasting, J., Hawley, S. 2010, Astrobiology, 10,751.

Shapley, H. 1954, AJ 59, 118.

Shkolnik, E.L., Barman, T.S. 2014, ApJ, 148, 64.

Snellen, I., de Kok, R., le Poole, R., et al. 2013, ApJ, 764, 182.

Snellen, I., de Kok, R., Birkby, J.L., et al. 2015, A&A, 576, 59.

Sparks, W.B., 1994, ApJ, 433, 19.

Sparks, W.B., Hough, J.H., Germer, T.A., Chen, F., et al., 2009, PNAS, 106, 7816-7821.

Sparks, W.B., Hough, J.H., Kolokolova, L., 2015, “Astrobiology” in Polarimetry of Stars and Planetary Systems, eds. L. Kolokolova, J. Hough, A.-C. Levasseur-Regourd, Cambridge University Press, ISBN 978-1-107-04390-9.

Stam, D.M., Hovenier, J.W., Waters, B.F.M. 2004, A&A, 428, 663.

Stam, D.M. 2008, A&A, 482, 989.

Turbet, M., Leconte, J., Selsis, F., Bolmont, E., Forget, F., Ribas, I., Raymond, S. N., & Anglada-Escud_e, G. 2016, A&A, 596, A112

Vida, K., Kovari, Z., Pal, A., Olah, K., Kriskovics, L. 2017, astro-ph/ 1703.10130

Wald, G. (1957) The Origin of Optical Activity. Annals of the New York Academy of Sciences, 69, 352–368.

Walker, G., Matthews, J., Kuschnig, R., et al. 2003, PASP, 115, 1023.

Yelle, R.V., Miller S., 2004, Jupiter. The planet, satellites and magnetosphere. Ed. Fran Bagenal, Cambridge University Press, ISBN 0-521-81808.

Zugger, M.E., Kasting, J.F., Williams, D.M., Kane, T.J., Philbrick, C.R. 2010, ApJ, 723, 1168.

Zuluaga et al. 2013, ApJ 770, 23

TABLE 1. Equations governing echo characteristic quantities: circular orbit

| Observable quantity | Equation* | Additional equations | Comments |
|---|---|---|---|
| Time delay $\Delta t$ | $\Delta t=(r/c)(1+\cos\Phi \sin i)$ | | |
| Radial velocity $v_r$ | $v_r=-v \sin \Phi \sin i$ | $v=\sqrt{\frac{GM}{r}}$ | Positive velocity is moving away from the observer |
| Positional shift $r_p$ | $r_p = r\,(1-\cos^2 \Phi \sin^2 i)^{1/2}$ | $x=-r \sin\Phi$<br>$y=r \cos\Phi \cos i$<br>$z=r \cos\Phi \sin i$ | x,y defines the sky-plane with line-of-nodes along x. z is distance along line of sight, z=0 in the sky plane. |
| Scattering angle $\zeta$ | $\cos \zeta = \cos \Phi \sin i$ | | |
| Polarization degree $p$ | $p=p_{max}(1-\cos^2\zeta)/(1+\cos^2\zeta)$ | | Rayleigh-like assumed |
| Polarization position angle $\xi$ | $\xi = -\tan^{-1}(\cos \Phi \cos i / \sin \Phi)$ | | $\xi=0$ implies polarization electric vector is parallel to the y-axis |

*For a fixed inclination i (i=0 is face-on), and orbital distance of the planet from the star, r, for a circular orbit, the equations are parameterized by the orbital phase angle $\Phi$ as the planet moves around the star, with $\Phi=0$ corresponding to superior conjunction (when the planet is behind the star). The orbital velocity $v=\sqrt{(GM/r)}$. In degrees, $\Phi=360\ t/T$, if t is time and T the orbital period.

TABLE 2. Equations governing echo characteristic quantities: elliptical orbit

| Observable quantity | Equation* | Additional equations | Comments |
|---|---|---|---|
| Time delay $\Delta t$ | $\Delta t=(r_e/c)(1+\cos\varphi \sin i)$ | $r_e=a(1-e^2)/(1+e \cos\theta)$ | $r_e$ is distance from focus (star) to planet |
| Radial velocity $v_r$ | $v_r=-v \cos \varphi_v \sin i$ | $\tan\beta= e \sin\theta/(1+e \cos\theta)$<br>$v=\sqrt{GM(2-\frac{r_e}{a})/a}$ | $\varphi_v$ is the direction of the velocity vector in the plane of the ellipse $\varphi_v=(\varphi_0+\theta+90-\beta)$ where $\beta$ is the flight path angle |
| Positional shift $r_p$ | $r_p = r\,(1-\cos^2\Phi \sin^2 i)^{1/2}$ | $x=-r_e \sin\varphi$<br>$y=r_e \cos\varphi \cos i$<br>$z=r_e \cos\varphi \sin i$ | |
| Scattering angle $\zeta$ | $\cos \zeta = -z/r_e$ | | |
| Polarization degree $p$ | $p=p_{max}(1-\cos^2\zeta)/(1+\cos^2\zeta)$ | | |
| Polarization position angle $\xi$ | $\xi = -\tan^{-1}(\cos \Phi \cos i / \sin \Phi)$ | | |

*For an elliptical orbit of eccentricity $e$ and major axis $a$, the equations are parameterized by the orbital phase angle $\varphi$ as the planet moves around the star, with $\varphi=0$ in the yz plane, and $\varphi=\varphi_0+\theta$ where $\theta$ is the true anomaly (angle from periapsis), and the angle of periapsis is $\varphi_0$.

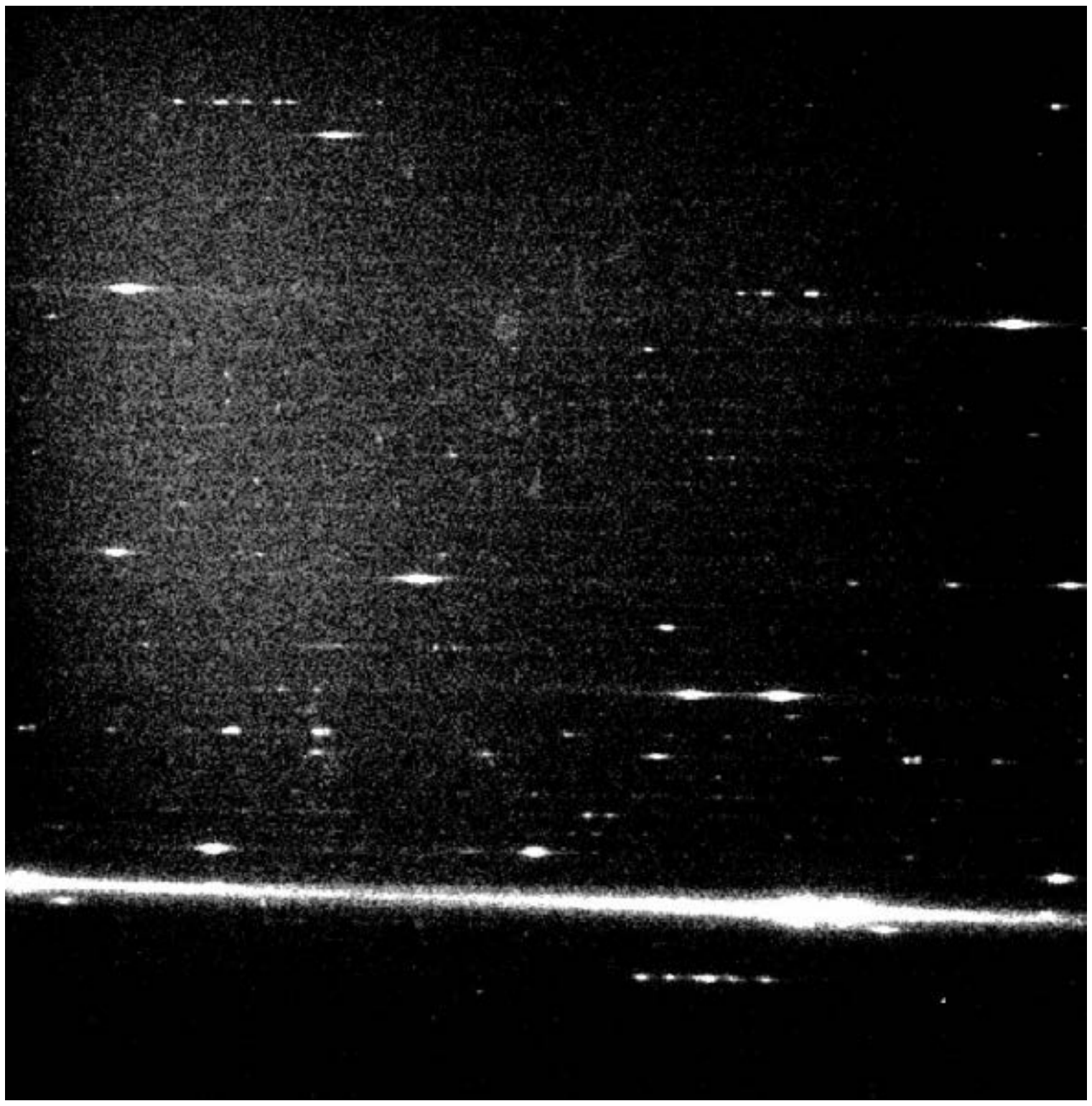

Fig. 1. Archival *HST/STIS* E140M spectrum of Proxima Cen, from 1123 Å (bottom) to 1710 Å (top). *All* of the many lines which can be seen flare dramatically. Lyα (near the bottom) exhibits smaller flaring amplitude, however it is subject to interstellar absorption and a potential geocoronal ingredient.

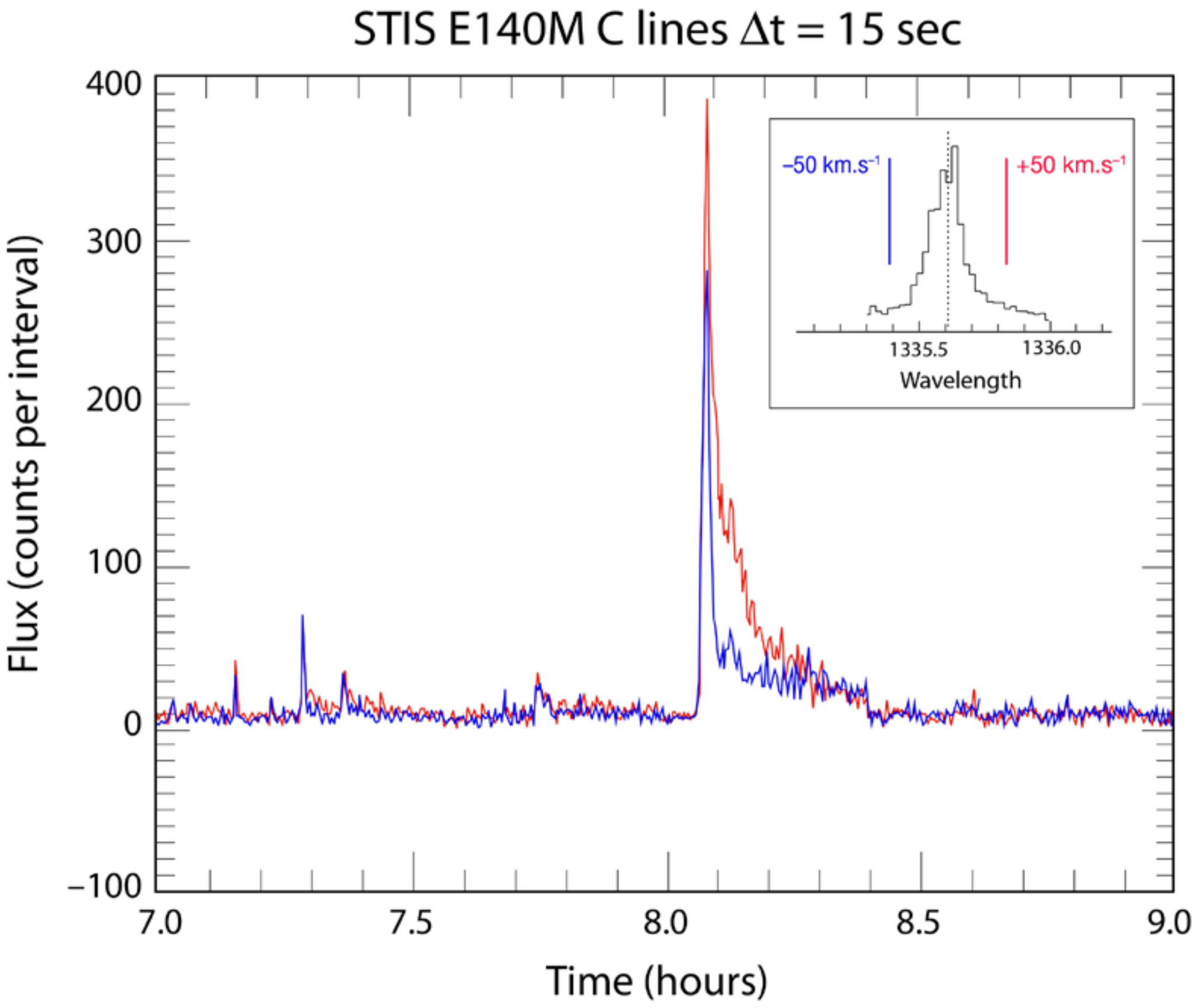

Fig. 2. *STIS* FUV time series showing the intensity through a bright flare of two of the many lines present – CIV (red) and CII (blue). The lines are bright, and exhibit a very strong impulsive phase, lasting ~10s seconds, ideal for light echo detection of planets in the habitable zone. The inset, upper right, shows the line profile for one of the CII lines and indicates a fiducial ±50 km $s^{-1}$ – a typical orbital velocity of the exoplanets of interest – showing that the Doppler shift can enhance contrast of planet to star by moving the planet's line emission well away from that of the star in wavelength.

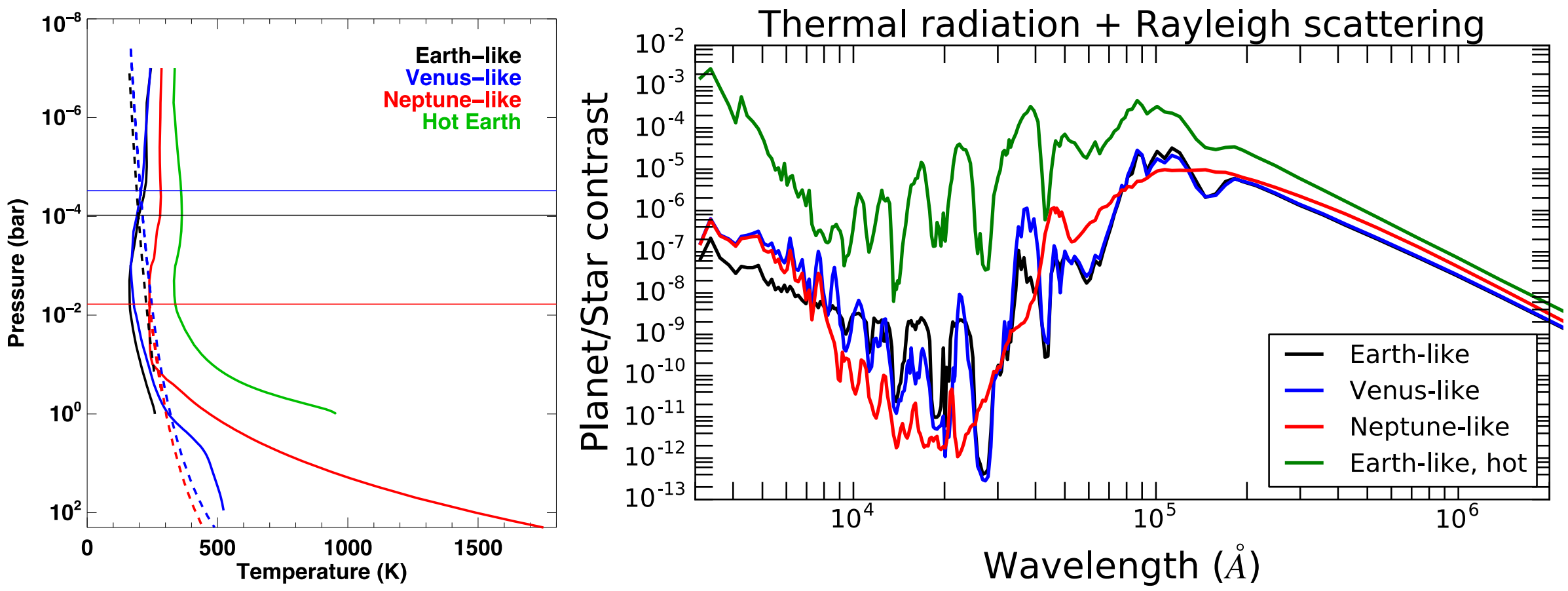

Fig. 3. Left: Pressure-temperature profiles for the four atmospheric composition scenarios described in the text for Proxima Cen b. The dashed lines show the water condensation curves, color coded after the corresponding models, and the horizontal lines show roughly the elevations of the cloud layers in each case. Right: Combined Rayleigh and thermal emission derived from the equilibrium structure. The contrast ratios emphasize the spectral windows most favorable for observations.

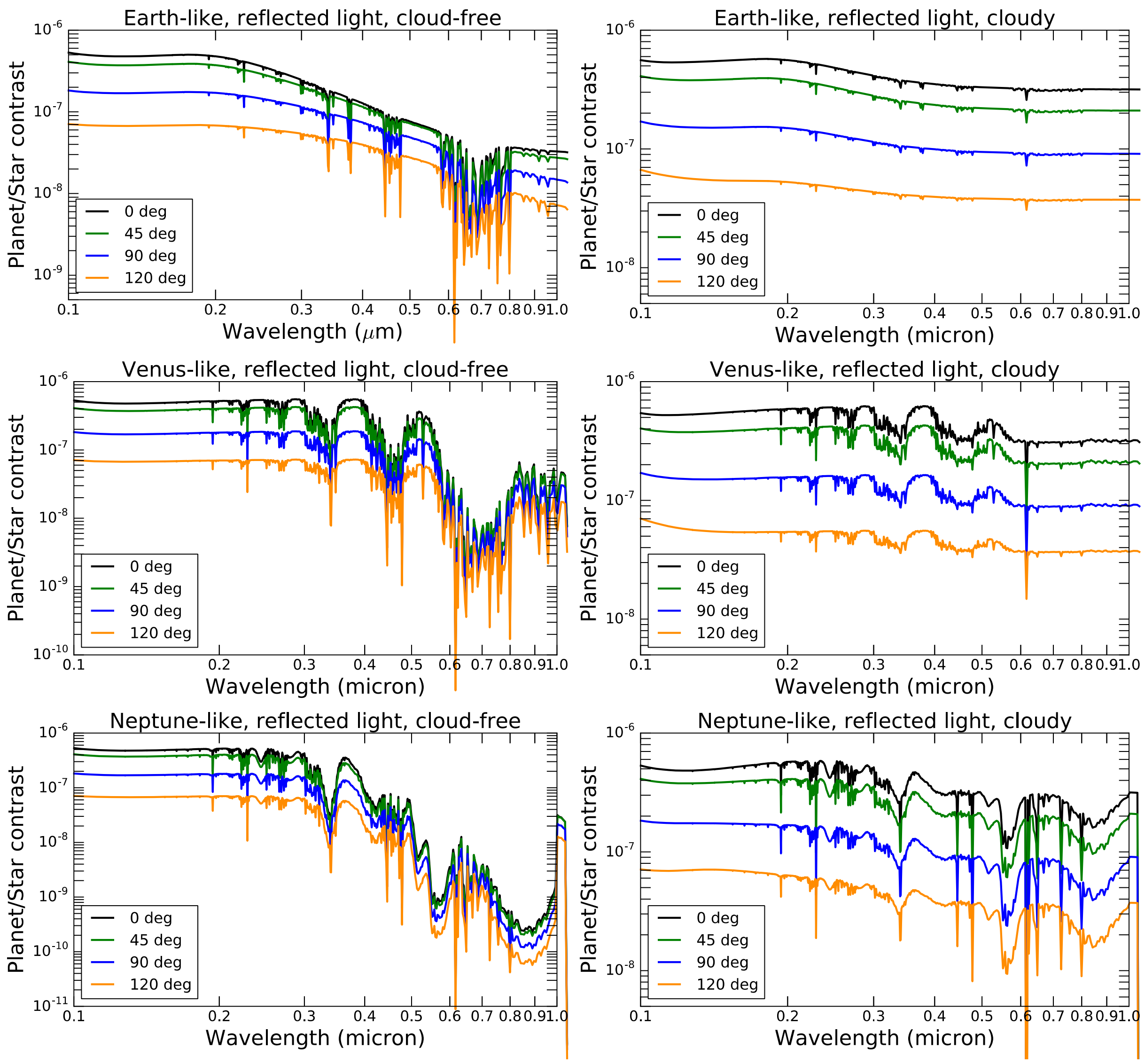

Fig. 4. Reflected light spectra as a function of phase angle for the three cold composition scenarios for Proxima Cen b. The clouds are assumed bright (albedo close to 1) and positioned at the altitude of water condensation. The cloudy models (right) show smaller absorption features, as expected from a smaller column of absorbers. Care must be taken, as absorption lines can obscure some transient features.

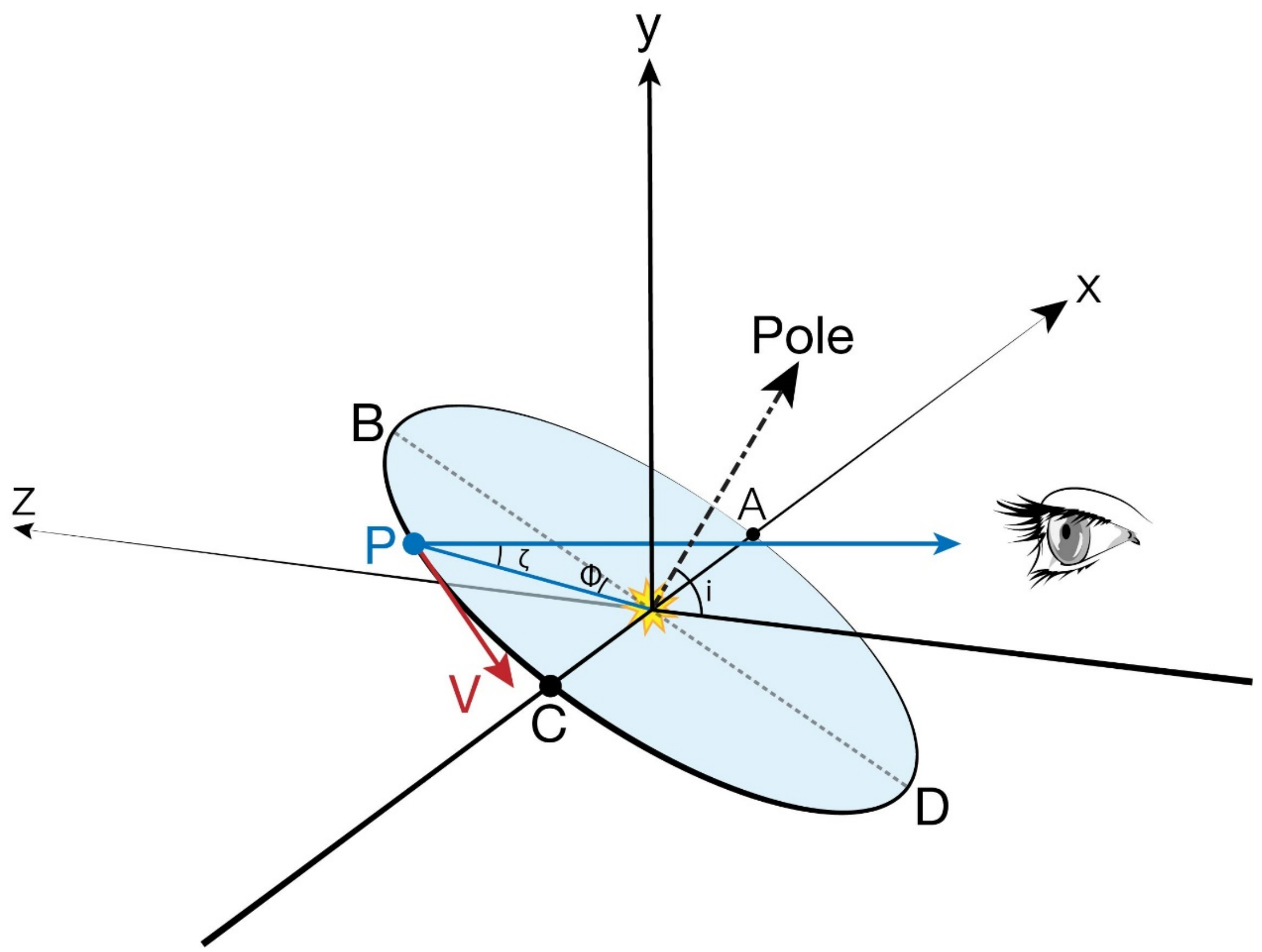

Fig. 5. Illustration of the geometry of an exoplanet's orbit around its host star, identifying the angles described in the text. The planet is at P, moving with velocity *v* about an orbit inclined at angle *i* to the line of sight (*z*). The sky plane is the *xy* plane.

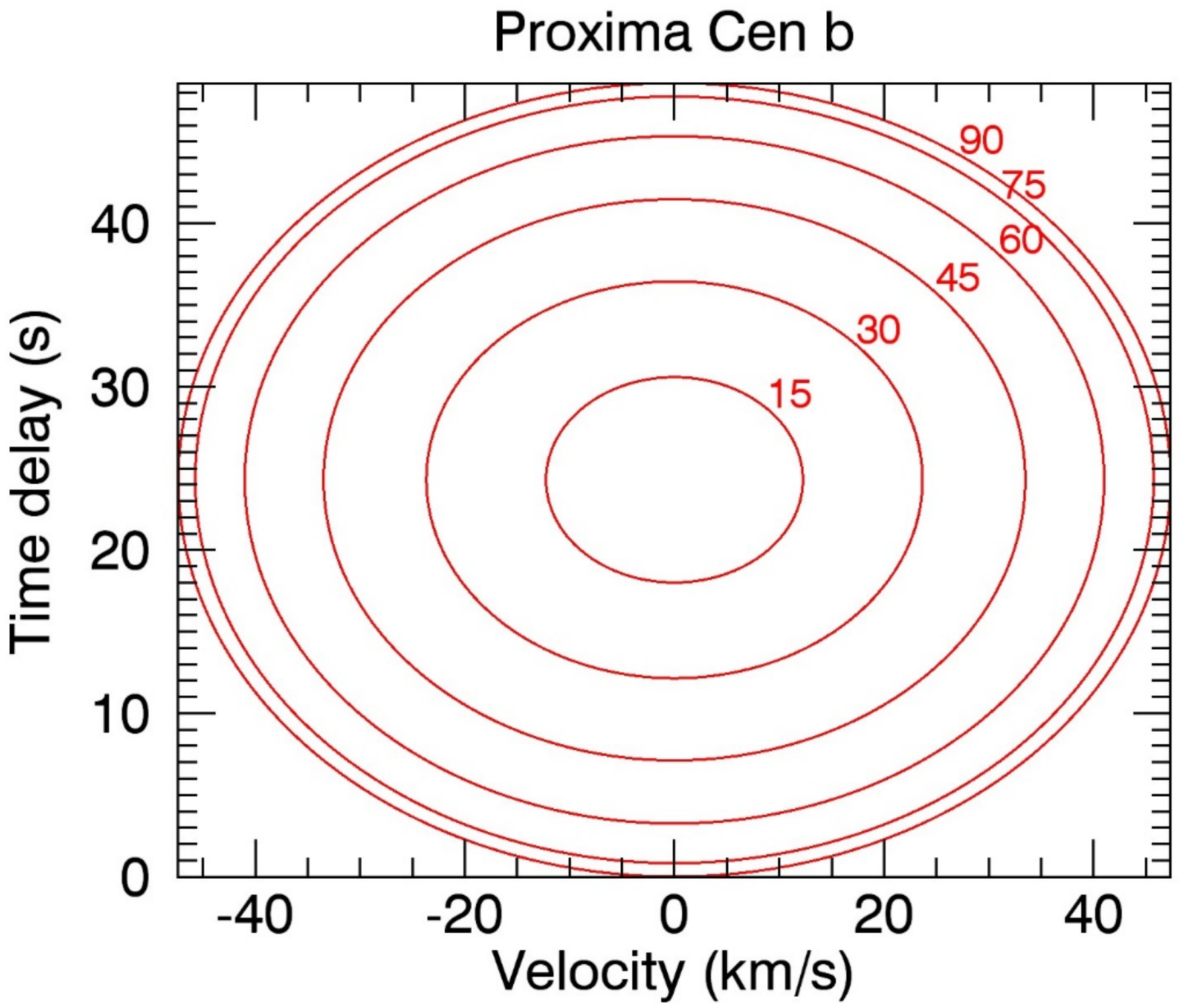

Fig. 6. For a circular orbit with parameters appropriate to the Proxima Cen b planet, the loci of velocity shift relative to the star versus time delay due to the extra distance traversed by scattered light from the planet, or light emitted by the planet due to the impact of the flare emission. Each contour is labelled by the inclination (degrees).

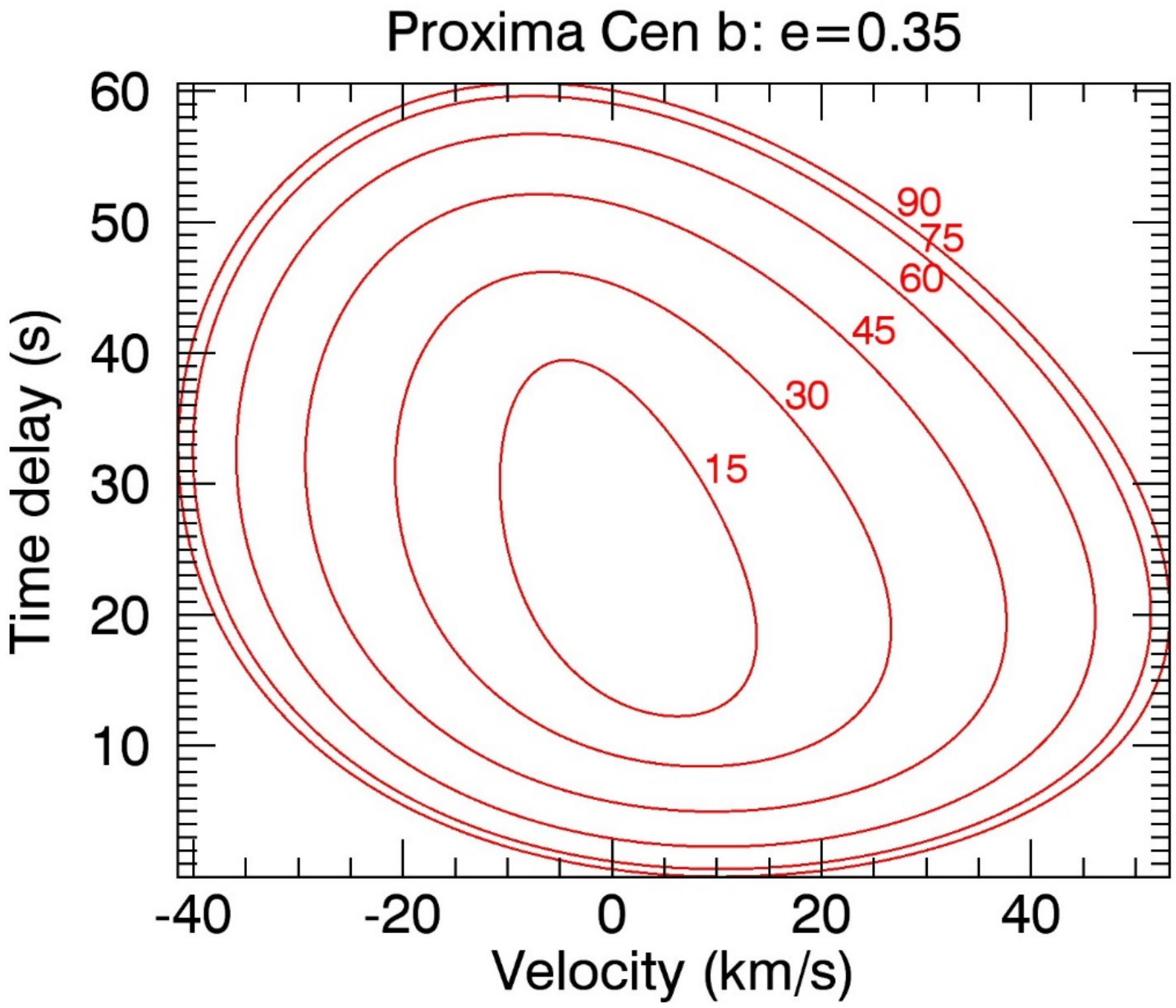

Fig. 7. For an elliptical orbit, the contours of Fig. 6 are distorted in a predictable fashion, according to the angle of periapsis, $\varphi_0$, and orbital eccentricity *e*. This figure is for an elliptical orbit with the upper limit to eccentricity *e*=0.35 appropriate to the Proxima Cen b planet. Contours are labelled by the inclination (degrees). The unknown angle of periapsis, $\varphi_0$ is taken to be 225° in this example.

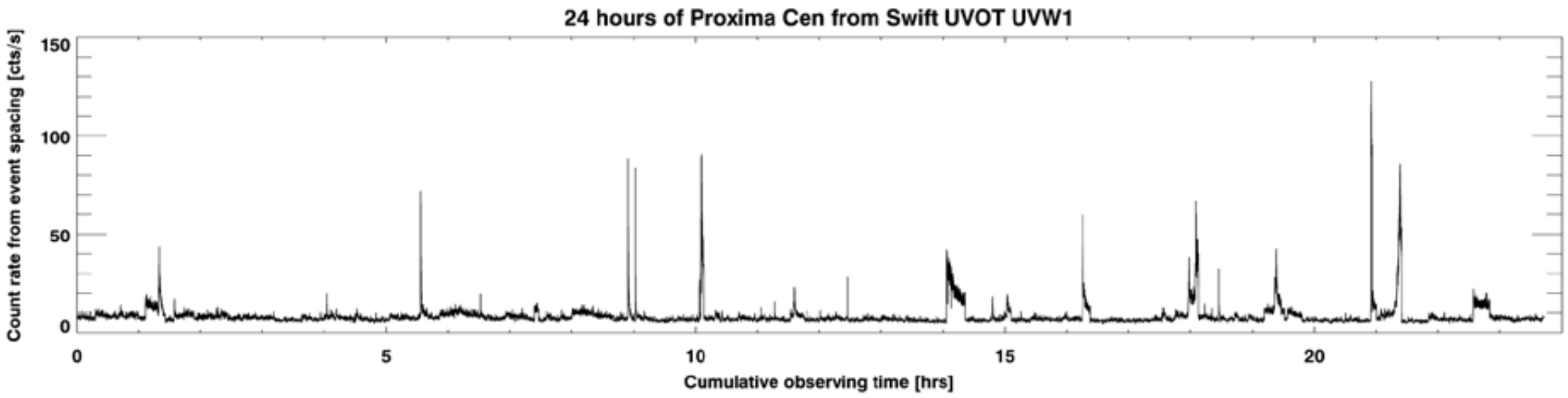

Fig. 8. Time series of broad band NUV observations of Proxima Cen from the SWIFT satellite, showing intense flaring activity.

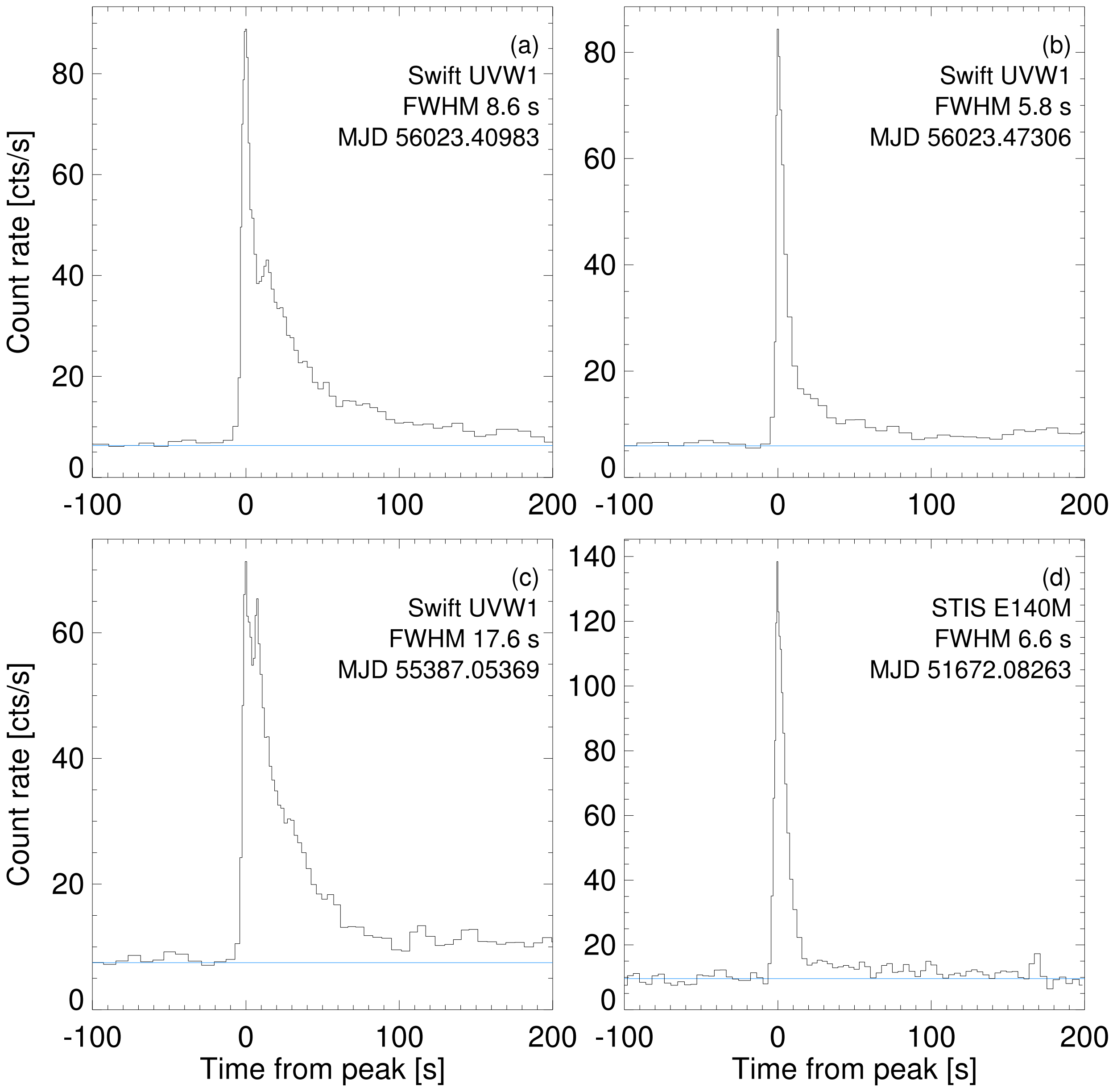

Fig. 9. Profiles of ultraviolet flares from Proxima Cen. Panels (a)-(c) show the 3 brightest flares seen by the Swift UVOT in 24 hours of observation using the *uvw1* filter. The FWHM in seconds and MJD of the peak are given. Panel (d) is a flare from the HST/STIS E140M FUV observation. These flares are frequently short compared to the light echo time delay for the Proxima Cen planet.

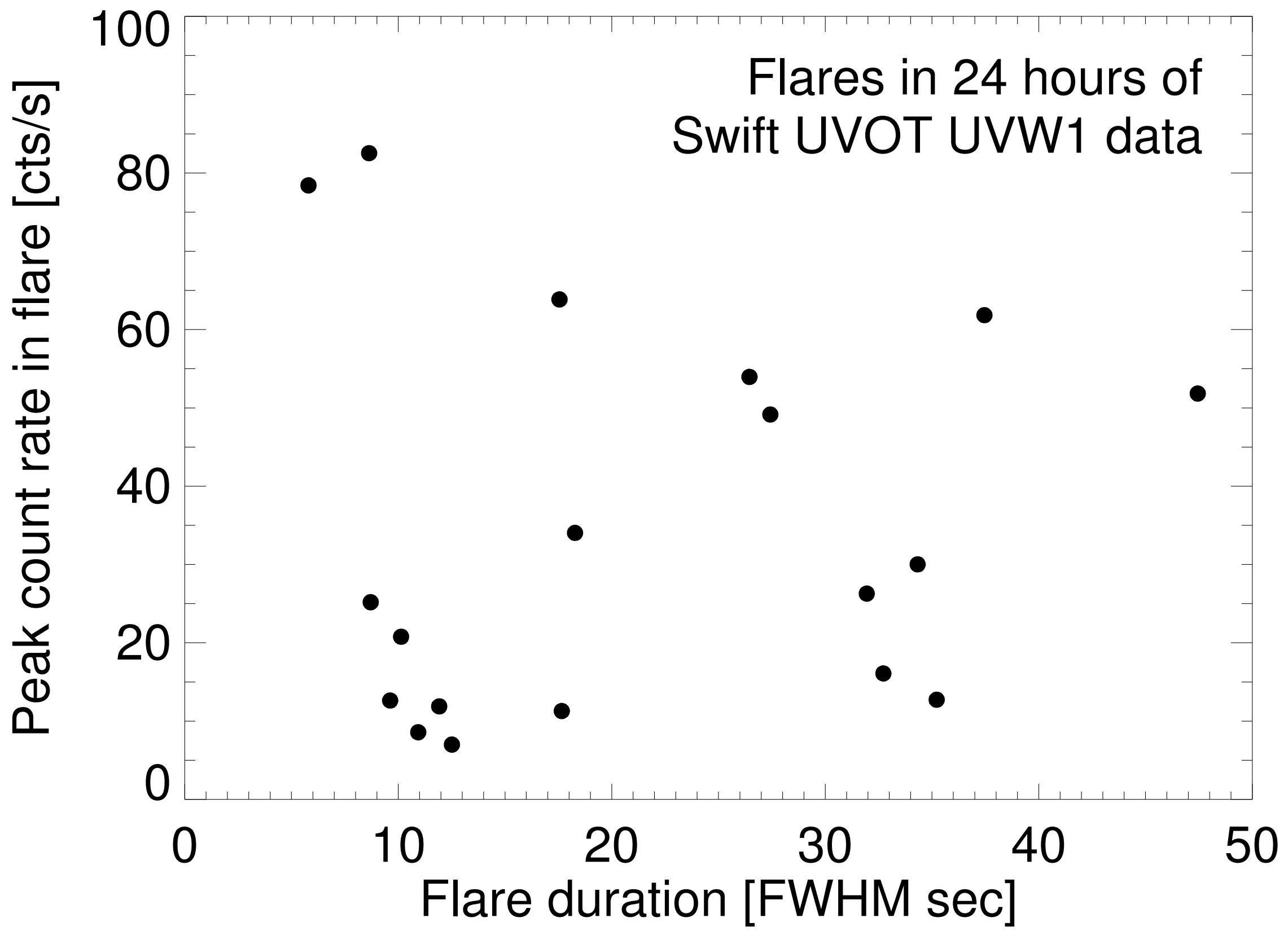

Fig. 10. Peak count rate in all 7-σ Swift UVOT *uvw1* flares as a function of the flare duration (FWHM). Note that the two strongest flares are also have the shortest durations. There is at most a weak correlation between flare strength and duration. The stronger flares that will be found in longer observations should also be short. Strong flares will give the best signal-to-noise for planet detection via light echoes.